\RequirePackage[table]{xcolor}%
\documentclass[AMA,STIX1COL]{WileyNJD-v2}

\articletype{Special Issue}%

\received{}
\revised{}
\accepted{}
        
\usepackage{url}
\usepackage[caption=false,font=footnotesize]{subfig}
\usepackage{amsmath}
\usepackage{colortbl}
\usepackage{color}

\begin{document}

\title{Optimization Policy for File Replica Placement in Fog Domains}

\author[1]{Carlos Guerrero*}
\author[1]{Isaac Lera}
\author[1]{Carlos Juiz}

\authormark{GUERRERO \textsc{et al}}

\address[1]{\orgdiv{Computer Science Department}, \orgname{University of Balearic Islands}, \orgaddress{\state{Palma}, \country{Spain}}}

\corres{*Carlos Guerrero, Building Anselm Turmeda. Crta. Valldemossa km 7.5. E07121 Illes Balears. Spain. \email{carlos.guerrero@uib.es}}


\abstract[Summary]{ Fog computing architectures distribute computational and storage resources along the continuum from cloud to things. It enables the execution of services or the storage of files closer to the users. The main objectives of fog computing domains are to reduce the user latency and the usage of the network. Availability is also an issue in fog architectures, where the topology of the network does not guarantee redundant links between the devices. Consequently, the definition of placement polices is a key challenge in this domain. We propose a placement policy for data replication to increase data availability with regard to other storage policies that only consider a single replica of the files. The system is modeled with complex weighted networks and topological features, such as centrality indices. Graph partition algorithms are evaluated to select the fog devices that store data replicas. Our approach is compared with other two placement policies: one that stores only one replica; and FogStore, which also stores file replicas but using a greedy approach (the shortest path). We analyze 22 experiments by simulation. The results show that our approach obtains the shortest latency times of the three policies and a file availability similar to FogStore. Additionally, our approach increase the network usage less than the FogStore.

}

\keywords{Fog computing, data placement, file replica management, complex networks, resource optimization.}

\maketitle

\section{Introduction}

In the last few years, there has been a significant increase in the number of applications developed for the Internet of Things (IoT). In this type of applications, it is considered that any device, however small, is able to connect to the Internet and monitor or control physical elements. These devices collect, process, and share data. As these environments have become popular, their needs for processing and storing data have increased rapidly, making it impossible to use the devices themselves to meet these storage and processing requirements.

A first solution to provide higher capacities to the IoT applications was to integrate cloud services by storing and processing the data of the IoT devices in the cloud providers and sending the results back to the IoT devices. This earlier solution allowed to develop applications that interact with real devices, thanks to the IoT paradigm, and that have unlimited storage and processing capabilities, thanks to the cloud. But important drawbacks emerged related to the high communication delays and the use of the network to send high quantities of data between the IoT devices and the cloud providers~\cite{8364042}. 

At the same time, the network communication devices have increased their computational capacities, and a new paradigm has been explored, the Fog Computing~\cite{Mahmud2018}. A fog architecture provides computational and storage capacities in the intermediate in-network devices. By this, the communication nodes are able to allocate services or to store data. This reduces the network latency of the applications, since the services are closer to the IoT devices. The data transmitted through the network is also reduced since it is stored closer to the IoT devices and only the processed and summarized data are finally sent to the cloud or to other IoT devices.

Important challenges and open research problems have emerged with the definition of fog architectures~\cite{8100873}. Data and service placement policies have an important role, since they are in charge of optimizing the performance of the system. Several service placement studies have established the first steps of future optimal solutions but, to the best of our knowledge, the efforts in the management of the placement of the data are lower. Data are achieving an important role in distributed systems due to the emergence of big data, and the definition of management and optimization policies is also an important challenge in smart and fog domains~\cite{8198795}.

We previously proposed a data placement policy for the optimization of data storage in fog devices with the objective of reducing the network usage~\cite{8364053}. This study proposed to store the data generated in the sensors as close as possible to the IoT devices by evenly distributing the distance between the storage device and the data sources. We studied the centrality indices of complex weighted networks as indicators to determine the fog device that is evenly close to the sensors. The results showed an important network usage decrease. The main contributions of the work were a first proposal for data placement in fog devices, and an optimization policy for the network usage based on the use of centrality indices.  Details of this previous work are summarized in Section~\ref{singledata}.

Fog architectures are environments where failures are not only likely but the common case due to, for example, device failures, network failures, battery drainage of mobile devices, shutdowns or restarts. The main requirements for distributed fog storage systems are fault tolerance, bandwidth consumption, and low latency~\cite{moysiadis2018}. Under these conditions, our previous work~\cite{8364053} lacks on ensuring the availability of the date because only one replica is available in the system. We propose to extend this previous work by creating replicas of the files to increase the fault tolerance. But having a higher number of replicas directly influences in the usage of the network bandwidth. The number of transmissions through the network is increased since a writing operation requires the update of a set of replicas, instead of only one file. Our proposal is addressed to increase as less as possible the network usage along the optimization of the latency and the improvement of the file availability.

The structure of the paper is as follows: Section~\ref{motivational} presents two examples to motivate our proposal;  Section~\ref{related} summarizes some related researches; we formally establish a model of our scenario in Section~\ref{system}; Section~\ref{data} explains the details of our data placement algorithm; Section~\ref{experimental} describes the experiments we carried out to evaluate our proposal, and Section~\ref{resultanalysis} presents and analyses the results of the experiments; finally, the conclusion and future works are included in Section~\ref{conclusion}.

\section{Motivational Scenario}
\label{motivational}

To motivate our work, we describe two simple examples based on a smart city with an intelligent transportation system and a factory with an IoT application for quality testing, which are typical scenarios of the use of fog computing~\cite{doi:10.1002/cpe.3485}. But our proposal can be applied to any IoT environment where data are stored and processed in disjoint sets~\cite{Bonomi2014,Dubey:2015:FDE:2818869.2818889}. 

Consider, as a first representative example, a city where the roads incorporate presence sensors that monitor the number of vehicles that are currently circulating. An IoT application is deployed in the smart environment that, by analyzing the current traffic status (number of vehicles), determines the best suitable routes for the vehicles. Digital road signs are placed on the roads to redirect the traffic flow in order to improve it.

The IoT application needs to process the data from the sensors by applying a big data algorithm, for example a MapReduce-based job, to determine the information to display in the road signs to control the traffic flow. These decisions are taken locally, and it is not necessary to process the data of all the city to determine the traffic flow in a given road or district of the city. Consider also that these decisions are taken by analyzing the current situation and historical data and, consequently, the data from a period of time need to be stored to be analyzed in the future.

An earlier solution, based on the use of cloud services that process the data and obtain the decisions, would require to transfer all the data from each sensor to the cloud provider. Fog computing emerged as a solution to reduce the latency and network usage in these cases, without compromising the computational and storage capacities of cloud services. An efficient data placement policy should determine the best fog devices, front a point of view of an optimized use of the network, to store the sensor data.   We addressed this issue in our preliminary work~\cite{8364053} by placing the file storage in a fog device equally distant to all the sensors of the roads, in order to reduce the network usage and the latency. In this preliminary work, we did not address a second important issue, data availability. Suppose that this storage location is in a device in a region of the city that is connected to the cloud only trough one single link. This link is a single point of failure between the storage and the services in the cloud that need to read the data. One solution to increase the data availability is to store several replicas in different regions of the network. We propose to extend our preliminary approach to increase availability by considering replicas and optimizing their placement to preserve low latencies and to increase the network usage as less as possible.

Consider, as a second example, a factory where electronic appliances are built. During the manufacturing process, several quality tests are taken by analyzing images of the electronic appliance. These quality tests are based on photographs taken in a set of consecutive phases of the manufacturing, but it is not necessary to analyze the photographs of all the phases for the consecutive tests. The photographs also need to be stored for future tests. Under these conditions, a pure cloud-based solution would require to transmit all the photographs to the cloud provider to be analyzed. An intermediate storage in fog devices could reduce the number of photographs transmitted to the cloud and  the image processing could be done in those fog devices, sending to the cloud only the results of the quality tests.   Note that only one single copy of the photographs does not guarantee a suitable availability for the future, since failures in fog devices are very likely. Replication is a solution to increase data availability, but replica allocation policies need to be defined to optimize this availability, along with the latency, and to reduce the increase of the network usage due to the update of the data in each replica.

\section{Related Work}
\label{related}

The problem of data placement and management has been widely studied previously for different types of distributed systems, such as data centers, data grids or the cloud~\cite{grace2014dynamic,hamrouni2016survey,malik2016performance,milani2016comprehensive, AlRuithe2018,10.1007/978-3-540-24688-6_11,10.1007/3-540-45414-4_4,Waibel2017}. But the characteristics of fog architecture have important differences with regard to these other storage distributed systems~\cite{moysiadis2018,DBLP:conf/fwc/MayerGSR17}: (a) the fog devices have limited resources in contrast to data center nodes which have high computational and storage capacities; (b) the fog devices are geographically distributed in wide areas through large-scale networks, in contrast with data center nodes which are located in the same geographic region; (c) the latency between nodes of a data center is negligible, contrary to fog domains where latency is a challenge; (d) nodes in data centers are usually connected via redundant links, in contrast to fog networks, where regions of the network can be connected to the rest of the network with a single link which, for example, could be a wireless one; (e) fog devices are very heterogeneous devices, in contrast to data center devices which usually have the same characteristics or very similar ones. These differences make necessary to re-evaluate traditionally solutions or the definition of new ones.

 Resource optimization techniques for fog computing have been mainly focused on the service placement problem. Previous proposals in the field of fog service placement are based on the use of integer linear programming~\cite{Velasquez2017,7511465,7676307, doi:10.1002/cpe.3975}, genetic algorithms~\cite{Skarlat2017}, greedy approaches~\cite{7110527}, Monte Carlo simulations~\cite{8014366,7919155}, and specific algorithm proposals~\cite{7249199,7987464}.

The number of studies addressing the problem of data placement in fog devices is more limited than for the case of service placement.  Although there is a review paper in the field of data management in fog computing~\cite{moysiadis2018}, it includes a very small number of papers that deal with data placement policies and with performance optimization. On the contrary, it includes a greater number of papers dealing with other challenges in fog data management.

In this last set of papers, Tang et al.~\cite{Tang:2015:HDF:2818869.2818898} presented a hierarchical distributed fog architecture to support the integration of a massive number of infrastructure components and services in future smart cities. Cecchinel et al.~\cite{6903302} defined a software architecture supporting the collection of sensor-based data in the context of the IoT. Another platform for IoT applications and Big Data analysis were defined by Cheng et al.~\cite{7207275} with the lessons learned from the experience in a real project of a smart city. Liu et al.~\cite{Liu2017} proposed a data collection approach where multiple nodes with similar readings formed a set, creating smaller sets near the data source and larger ones in the sets far from the data source. They used a multi-representative re-fusion approach. 

Some optimization policies in fog architecture have been defined, not for data management but for data caching. For example, Vural et al.~\cite{6883811} studied in-network caching of IoT data with a model based on the trade-off between multihop communication costs and the freshness of a transient data item. Optimization policies to address the definition of cloudlets in the edge devices of the network have been also proposed, for example, in the work of Fan and Ansari~\cite{7996722}.

To the best of our knowledge, the work of Mayer et al.~\cite{DBLP:conf/fwc/MayerGSR17} and Gupta and Ramachandran~\cite{Gupta:2018:FGK:3210284.3210297} is the most similar proposal for replica placement, since it deals with replica placement and performance optimization, and, additionaly, with data consistency. They proposed, similarly to our case, the partition of the fog architecture into groups (called failure groups) and these groups are used by the placement algorithm to decide the replica placements. This proposal has three main limitations: the failure groups are defined by human experts who knows the conditions of the network topology; only one data source is considered for each data type (or file); the partition of the network is not defined by considering the placement of the data sources. In contrast, our proposal is completely autonomous and non-human assisted, considers several sources for the data types, and takes into account the data sources for the partition of the network, i.e., the partition of the network is different for each data type.

A set of works have previously applied complex network algorithms to the field of distributed systems.   The use of topological metrics from the field of complex networks was proposed for the management of VM placement in data centers~\cite{Filiposka2015}, or for load balancing in cloud federations~\cite{5538385}. Later studies have also proposed to use complex network theory in the field of fog computing for the optimization of application placement and migration~\cite{8377095}, data storage~\cite{8364053}, or fog architecture organization~\cite{Guerrero2018ontheinfluence}. Although those preliminary studies, there is still room for wider application of those techniques in fog architectures.

Additionally, we have also previously investigated in the field of data replica placement for other architectures and the use of complex networks in other problem domains. For example, we have investigated the placement of data replicas in virtualized Hadoop architectures~\cite{guerrero2018multiobjective}. We proposed to address this problem by simultaneously managing, first, the allocation of the virtual machines (that implemented the DataNodes) into the physical machines and, second, the placement of the file replicas into the DataNodes. We proposed to use genetic algorithms for solving this optimization problem. The features of the data placement in a fog architecture completely differs from the data placement in a virtualized cloud. 
 
We have also investigated the placement of services in fog domains in two previous works: we proposed the use of a greedy decentralized algorithm which allocated the most requested services closer to the users~\cite{Guerrero2018lightweight}; and we proposed the placement of services in fog devices by mapping the partitions of the application graphs (using the graph transitive closures) with the partitions of the device organization (using complex network metrics)~\cite{8588297}. The problem domain of service placement mainly differs from the data placement in the consequences of scaling the replicas. Scaling up the number of data replicas increases the network usage due to the increase of writing operation that need to update all the replicas. On the contrary, scaling up the number of services reduces the network usage because the user request are executed in closer devices. Thus, the features of the problem, the experiment design and the analyzed metrics of data replica placement are different with regard to the service allocation problem.

We have also used complex network metrics in the field of fog devices to determine the formation of device clusters, or fog colonies~\cite{Guerrero2018ontheinfluence}. This work studied the intra and inter network distance of the fog colonies as the number and size of colonies are changed. 

In summary, the novelty of our approach consists in: increasing the file availability by storing several replicas; using complex network topological metrics for the partition of a fog architecture into regions; taking into account the location of the data sources for the partition of the network; allowing multiple data sources for each file; using centrality indices to determine the placement of file replicas inside the regions; keeping low the file reading and writing latencies; increasing as less as possible the network usage due to the update of each replica.

\section{System model}
\label{system}

The physical elements in our architecture are: (a) fog devices, network devices with storage and processing capacities; (b) sensors, the elements that generate data to be stored and analyzed by the IoT applications; (c) gateways, the network devices where the sensors are connected to; (d) the cloud provider, that is defined as an element with unlimited computational and storage capacities.

The logical elements are: (a) files, that store the data generated in the sensors; (b) file replica, each of the identical copies of a file and that are stored in different fog devices; (c) data producer type, the data generated in a set of sensors and that is stored together in the same file; (d) data consumer type, the set of devices that read the same file stored in the fog domain.

\begin{figure*}[!t]
	\centering
	\subfloat[Graphical representation of the example.]{\includegraphics[width=3.5in]{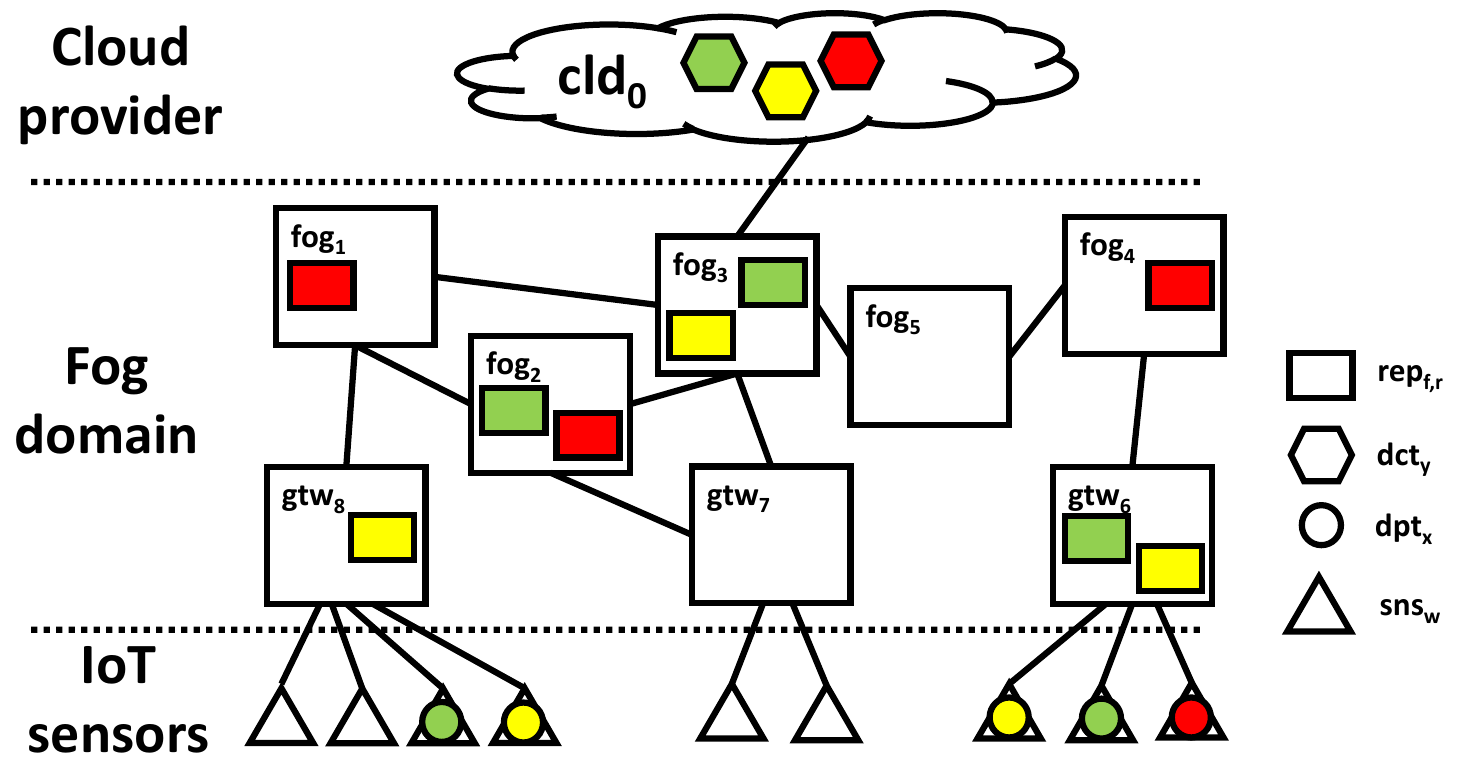}%
		\label{fig:fogarchitecture}}
	\hfil
	\subfloat[Model representation of the infrascturcture of the example (physical layer).]{\includegraphics[width=2.0in]{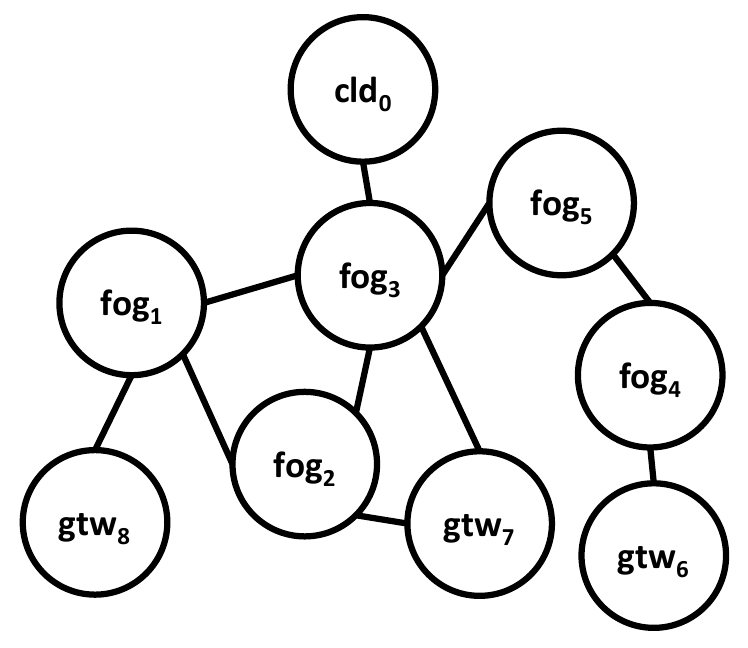}%
	\label{fig:fogarchitecturemodel}}
\hfil
	\subfloat[Model representation of the fle replicas of the example (logical layer).]{\includegraphics[width=3.5in]{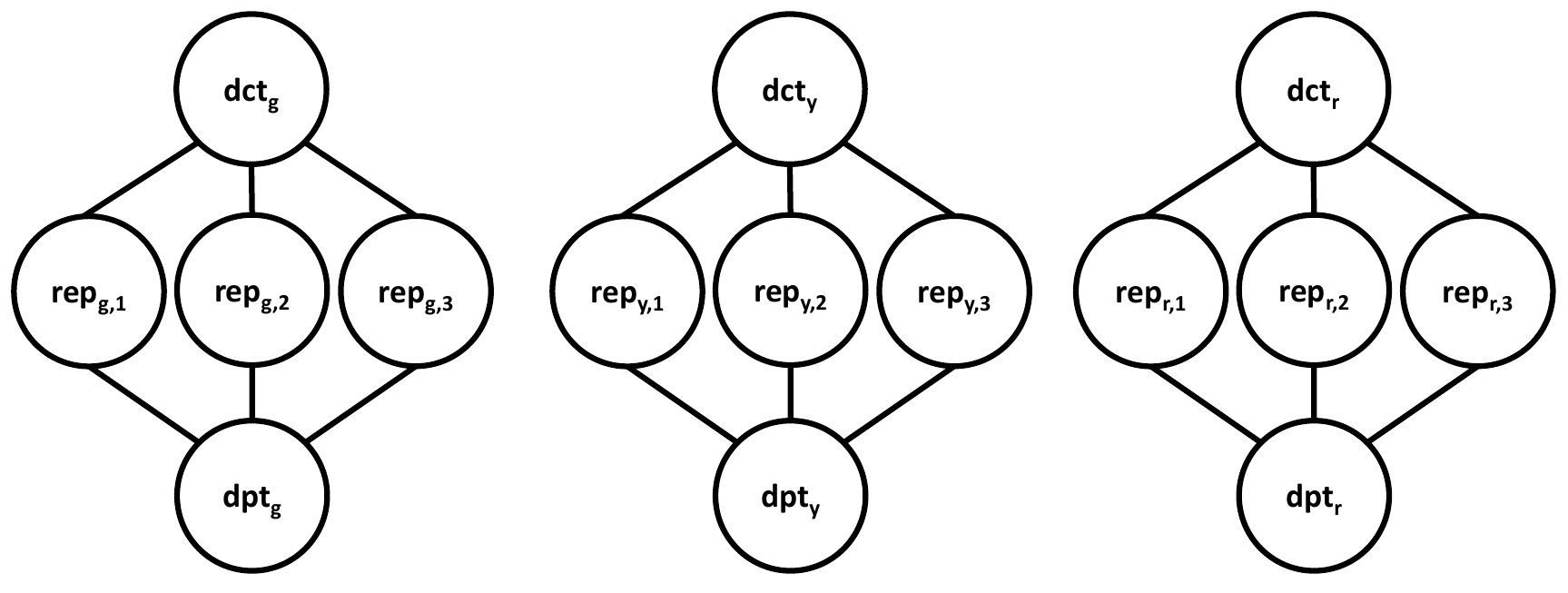}%
	\label{fig:filemodel}}
	\hfil
	\caption{Example of a fog domain with storage capacities.}
	\label{fig:fogexample}
\end{figure*}

Figure~\ref{fig:fogarchitecture} shows a graphical representation of those elements, using a cloud symbol to represent the cloud provider, a square for the fog devices and gateways, a triangle for the sensors, a colored rectangle for the file replicas, a colored hexagon for the data consumer types, and a colored circle for the data producer types.

We have used complex networks to model the components of the system. Two model layers are defined, one for the physical components and other for the logical ones. We call Fog File Replica Placement Problem (FFRPP) to the mapping between both model layers to determine the placement of the file replicas into the fog devices.

The physical layer is defined by the  graph structure of the communication network topology, $G = (DEV,LNK)$. $DEV$ is the set of all the devices $dev_i$ and $LNK$ is the set of all the network links $lnk_j$ between two devices.   $|DEV|$ denotes the total number of devices in the system.   A network link is defined only when a direct connection exists between two devices. The four elements in the physical layer are identified as follows: $fog_u$ for the fog devices; $gtw_v$ for the gateways; $sns_w$ for the sensors; and $cld_c$ for the cloud provider. Gateways and cloud providers are just specific types of fog devices to indicate that they have, respectively, unlimited resources, or sensors connected to. All of these three elements are able to store and allocate files.  We have considered only one cloud provider for our particular solution proposal, but the use of multiple clouds could be easily generalized. The fog devices are characterized with their storage capacities ($dataCap_u$). The network links are characterized with their network propagation time ($netPr_j$) and bandwidth ($netBdw_j$), which define the latency for each transmitted package. The infrastructure of the example of Figure~\ref{fig:fogarchitecture} is modeled and represented as a graph in Figure~\ref{fig:fogarchitecturemodel}. This graph is the model of the physical layer, and it is formed by 9 nodes that represent 3 gateways ($gtw_6$, $gtw_7$, and $gtw_8$ ), 5 intermediate fog devices ($fog_1$, $fog_2$, $fog_3$, $fog_4$, and $fog_5$) and the cloud provider ($cld_0$).

The logical layer is also defined as a graph structure, which represents the files in the system.   $|FILE|$ denotes the total number of files in the system.   A file $file_f$ is defined by its replicas and its producers and consumer data types. The nodes of the graph represent those three components of a file: the replicas of the file ($rep_{f,r}$), which are connected to two special nodes, the node that represents the data producer type, $dpt_f$, and the one that represents the data consumer type, $dct_f$. Consequently, sensors are characterized by the data type they produce ($dpt_f$), and the data produced by all the sensors of the same type are stored in the same file. Additionally, data producer types are characterized: by the storage size that needs to be provisioned in the fog devices that would store the replicas of file, $dataReq_f$; by the data generation rates ($writeRate_f$) that is the summation of the data generation rates of the sensors with the same data type; and, finally, by the data packet size that is the size of the packets transmitted from the sensor to the fog device ($writePacketSize_f$). In the same way, the data consumer types are also characterized with the data reading rates ($readRate_f$) and sizes ($readPacketSize_f$), that obviously are different to the values of the producer because the data can be processed and summarized in the fog devices. The replication factor of a file (the number of copies of the file in the system) determines the number of $rep_{f,r}$ nodes of each file. The three graphs in Figure~\ref{fig:filemodel} represent the three files (replicas and data types) in the example of Figure~\ref{fig:fogarchitecture}. All the files have a replication factor of 3, thus, the number of nodes in each file graph is 5: 3 for the replicas ($rep_{f,1}$, $rep_{f,2}$, and $rep_{f,3}$) and 2 more for the data consumer ($dct_f$) and the data producer types ($dpt_f$).

\begin{figure*}[!t]
\centering
\subfloat[Static mapping between physical layer (gray background) and logical layer]{\includegraphics[width=3.5in]{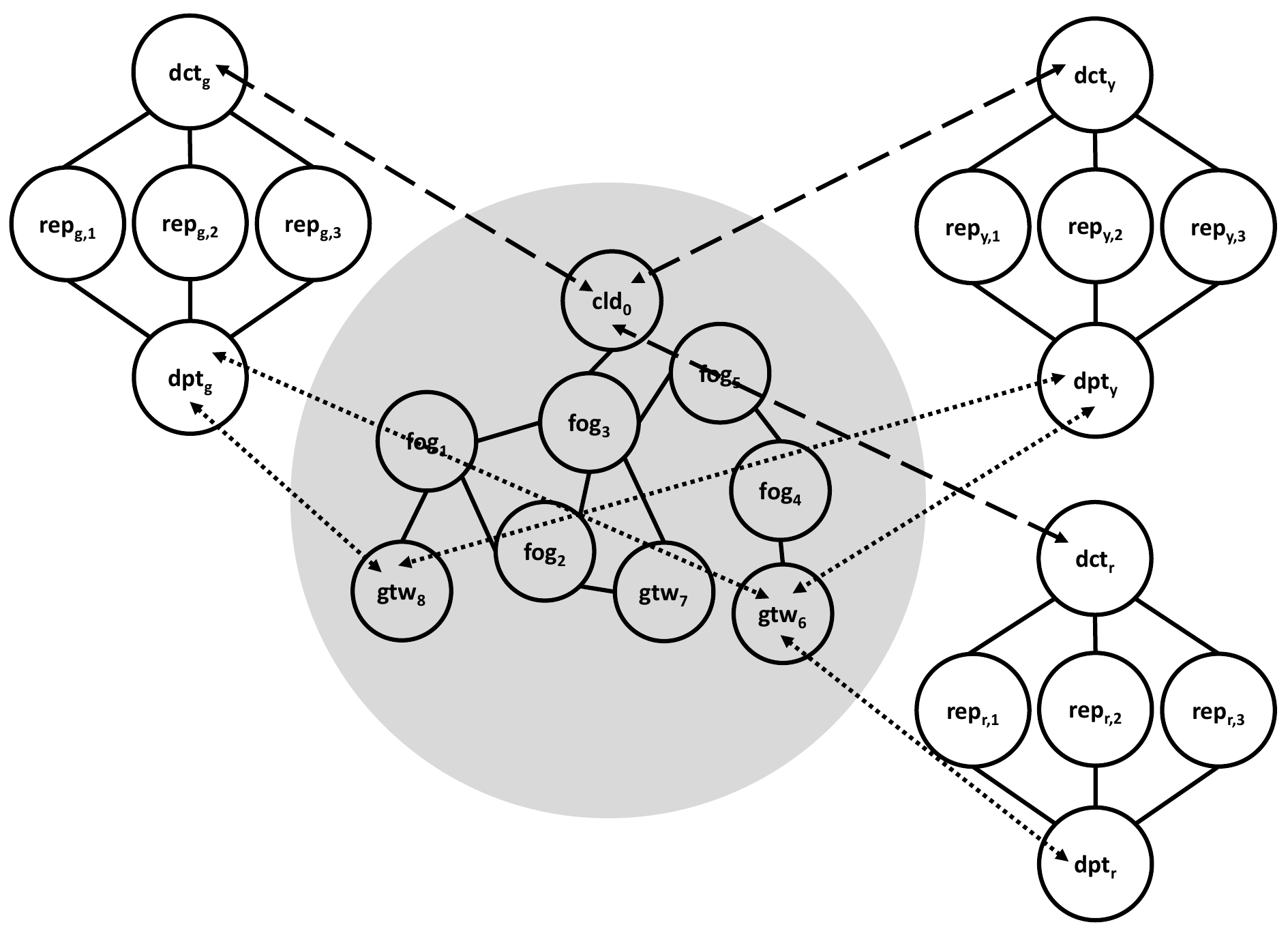}%
\label{fig:staticmapping}}
\hfil
\subfloat[Dynamic mapping between physical layer (gray background) and logical layer]{\includegraphics[width=3.5in]{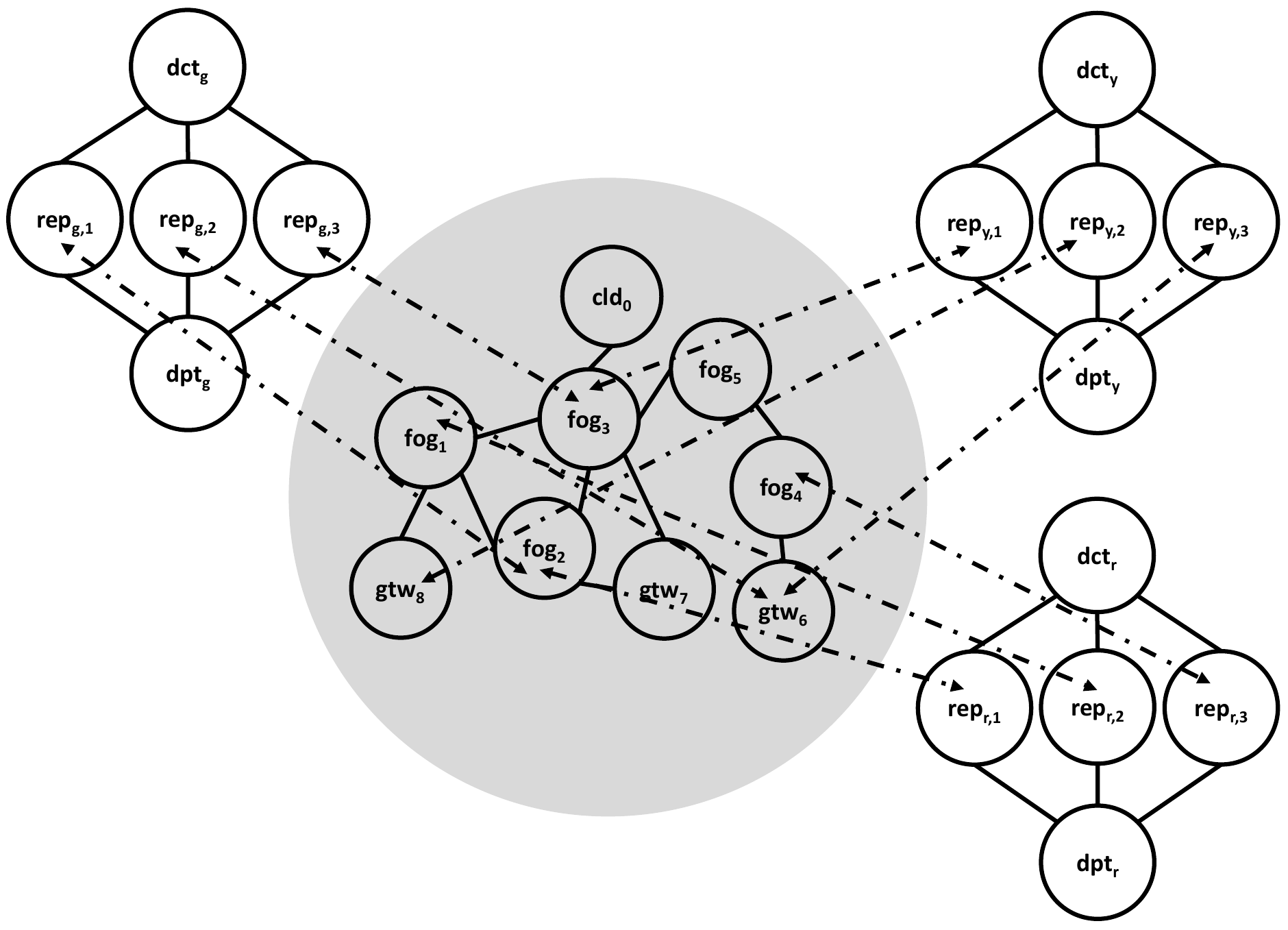}%
\label{fig:dynamicmapping}}
\caption{Model representation of the example of Figure~\ref{fig:fogarchitecture}.}
\label{fig:modelexample}
\end{figure*}

The use of this representation for the logical layer is general enough to also represent file chunks (using more than one graph node to each replica), or even file chain flows where the data consumer type of a file is the data producer type of the following file. Despite this, our proposed solution is addressed to optimize the constrained case of fixed number of replicas (three), with non-split files, and single data flows where sensors are the data producer types and the cloud provider is the unique data consumer type for all the files.

The Fog File Replica Placement Problem (FFRPP) addresses the decision about the fog devices that store the file replicas, i.e, the mapping between file replicas and fog devices. This results in mapping the physical and logical layers of our model. Additionally, our model also maps some elements that are not included in the FFRPP but that define the conditions of the domain scenario: the data producer types with the gateways where the sensors are connected to, and the data consumer types with the cloud provider (because of the constraint on our proposed solution). Figure~\ref{fig:staticmapping} shows the mapping between the models of the examples in Figure~\ref{fig:fogexample}. For instance, it is observed that all the data consumers ($dct_f$) are mapped to the cloud provider ($cld_0$).

As we explained before, the logical model could be extended for a more general scenario by including relationships in the logical layer between data consumer type nodes $dct_f$ and data producer type nodes ($dpt_f$). In any of both cases, those relationships are not determined in the optimization process of the FFRPP, and they are defined by the system conditions (for example, the physical places where the sensors have been deployed and connected). Thus, we have considered two different mappings between the physical and the logical layers: the static mapping, that represents those previous relationships (Figure~\ref{fig:staticmapping}); and the dynamic mapping, that represents the placement of the replicas in the fog devices (Figure~\ref{fig:dynamicmapping} shows the file allocation of the previous examples). Note that a solution proposal for the FFRPP only addresses this second case, the dynamic mapping.

Fog computing was proposed to solve the problems of latency and network usage in cloud-based Internet of Things applications~\cite{doi:10.1002/ett.3307}. Consequently, any new proposal in the domain of fog computing should also deals with this concerns. Additionally, one of the main problems in distributed file systems is data availability~\cite{969727}. 

The solution to the FFRPP is the mapping between data replicas and fog devices. A fog device that allocates a file replica is in charge of storing all the data generated in the sensors of the given type. We use a binary matrix $P$ to represent that a file, $file_f$ is stored in a specific fog device, $fog_u$. The matrix is defined as:

\begin{equation}
	P =  \begin{cases} 
		p_{fu}= 1 &  \text{if a replica of}\ file_f\ \text{is stored in}\ fog_u  \\
		p_{fu}= 0 & \text{otherwise}
	\end{cases}
\end{equation}

We have constrained the problem to consider a static and general replication level of the file, fixed in tree copies. Thus, each line of matrix $P$ includes three elements with value 1:

\begin{equation}
\label{const1}
	\sum_{u=1}^{|DEV|} p_{fu} = 3, \forall\ f=(1,\ldots,|FILE|)
\end{equation}

A second constraint is also defined. The summation of the storage requirements (provisioned size) of all the replicas allocated in a fog device should be smaller than the total storage capacity of the device:

\begin{equation}
\label{const2}
	\sum_{f=1}^{|FILE|} p_{fu}\times  dataReq_f \leq dataCap_u, \forall\ u=(1,\ldots,|DEV|)
\end{equation}

The first optimization objective of our algorithm is file availability. We measure the availability in terms of the writing and reading operations. We consider the file is available for reading if the cloud provider can communicate with at least one of the file replicas. The mathematical expression for file availability for reading operations is:

\begin{equation}
filAvl_{read} = \frac{\sum_{file_f}^{|FILE|} max_{u}^{|FOG|} \left( reachable(cld_0,fog_u), | p_{yu}=1 \right)}{|FILE|}
\end{equation}

where:

\begin{equation}
reachable(u, u') =  \begin{cases} 
1 &  \text{if}\ \exists\ a\ network\ path\ between\ u\ and\ u'  \\
0 &  \text{if}\ \nexists\ a\ network\ path\ between\ u\ and\ u'  \\
\end{cases}
\end{equation}

This formula just calculates the percentage of files that can be accessed from the cloud provider to at least one of their replicas in the system.

In the case of writing operations, we consider that a file is available if all the sensors that generate the type of data of the given file are able to reach at least one replica of the file. We suppose that some consistency policy exists in the system to be able to update the replicas that have been modified during the period of times that not all the replicas are reachable. Thus, for our case, it is enough to have access to just one of the replicas. 

The definition of a consistency policy is out of the scope of our work and the integration with any approach keeps as an open research challenge for future works. But our placement algorithm could be integrated in some previous proposal, such as the proposal of Mayer et al. called FogStore~\cite{DBLP:conf/fwc/MayerGSR17,Gupta:2018:FGK:3210284.3210297}, keeping the same consistency algorithm that it included, but replacing the placement algorithm.

The file availability in writing operations is calculated as:

\begin{equation}
filAvl_{write} = \frac{\sum_{file_f}^{|FILE|} min_{w}^{sns_w \in dpt_f}\left[ max_{u}^{|FOG|} \left( reachable(cld_0,fog_u), | p_{yu}=1 \wedge sns_w \in gtw_v \right) \right]}{|FILE|}
\end{equation}

This formula just calculates the percentage of files whose all the sensors are able to reach at least one of the file replicas. If there is a sensor that is not able to reach any of the replicas, we consider that the whole file is not able for writing since the data of this sensor would not be stored anywhere. 

Our second optimization objective, the file access latency, is also measured in terms of the writing and reading operations. In the case of the readings, the latency is defined as the minimum of the shortest latency paths between the cloud provider and the file replicas, since we consider that a data selection policy would select to access to the closer replica. The details of this selection policy are out of the scope of the paper, but there are many research works that have analyzed the replica selection problem in other domains~\cite{KINGSYGRACE20142099}. The average of the reading access latencies for all the files in the system is calculated as follows:

\begin{equation}
accLat_{read} = \frac{\sum_{f}^{|FILE|}min_{u}^{|FOG|} \left(  shortestPathDist(cld_0, fog_u), \ | \ p_{fu}=1 \right)}{|FILE|}
\end{equation}

where  $shortestPathDist(cld_0, fog_u)$ is the latency of the shortest path between a fog device and the cloud provider, measured as the summation of all the latencies of each single link in the shortest path ($netLat_j$). 

On the contrary, a writing operation is performed in all the replicas. To illustrate the range of writing latencies, we consider the averages of the maximum and minimum values of the shortest latency paths between the sensors and the replicas, since the writing operations are performed in parallel. The average maximum writing latency is calculated as follows:

\begin{equation}
accLat^{max}_{write} = \frac{\sum_{f}^{|FILE|}\sum^{sns_w \in dpt_f}max_{u}^{|FOG|} \left(  shortestPathDist(sns_w, fog_u), \ | \ p_{fu}=1 \right)}{\sum_{f}^{|FILE|}\sum^{sns_w \in dpt_f} 1.0}
\end{equation}

The minimum latency is calculated in the same way but considering the $min()$ function instead of the $max()$.

Finally, the network usage is also an important concern in fog domains. We also separately analyze the network usage in terms of writings and readings. We measure the network usage with the number of messages that are transmitted across the communication network. This metric considers that a message is transmitted for each single network link that forms part of the network path between two nodes. For example, if sensor $sns_w$ transmits one packet from $gtw_v$ to $fog_u$ through the network path $gtw_u \rightarrow fog_{u'}  \rightarrow fog_{u''}  \rightarrow fog_{u}$ , the number of transmitted packets is three. The total number of message transmissions for writings between each pair of sensor and replica is calculated as follows:

\begin{equation}
msgTrns_{write} = \sum_{f}^{|FILE|}\sum_{u}^{|FOG|} p_{fu} \times \sum^{sns_w \in dpt_f} hopCount(sns_w, fog_u)
\end{equation}

where $hopCount(sns_w, fog_u)$ is te number of link connections in the shortest path between the gateway of the sensor and the fog device where the data are allocated. 

The total number of message transmissions for readings between a file and the cloud provider, also considering that only the closer replica is accessed by the cloud, is calculated as follows:

\begin{equation}
msgTrns_{read} = \sum_{f}^{|FILE|}min_{u}^{|FOG|} \left(  hopCount(cld_0, fog_u), \ | \ p_{fu}=1 \right)
\end{equation}

In summary, our solutions is addressed to determine matrix $P$ by minimizing $filAvl_{read}$,  $filAvl_{write}$,  $accLat_{read}$, $accLat^{max}_{write}$, $accLat^{min}_{write}$,  $msgTrns_{write}$, and 
$msgTrns_{read}$ considering the constraints in Equations~\eqref{const1} and~\eqref{const2}. 

\section{Data Placement Algorithm}
\label{data}

This section describes the new proposed algorithm to allow the placement of replicas of the data in a fog architecture (Section~\ref{replicadata}), and a summary of our previous proposal~\cite{8364053} for the placement of files that are not replicated (Section~\ref{singledata}).

\subsection{Replica-aware Data Placement}
\label{replicadata}
We propose to optimize the FFRPP implementing a similar solution to the rack-awareness policy from Hadoop File System (HDFS)~\cite{guerrero2018multiobjective,Guerrero2018}. HDFS is usually deployed in data centers where the nodes are organized in racks and its replica placement policy stores at least two replicas in nodes from different racks to guarantee the file availability if a rack gets down.

In fog domains, the nodes or devices are not organized in racks since they are geographically distributed and more heterogeneous than in a data center. But fails in a device or link could result in the unreacheability of some regions of the network. Consequently, it is necessary to consider the topological distribution of the fog devices, the data generators, and the cloud provider, and the distances between them to improve file availability. Complex networks are suitable modeling tools when the topological structure of a domain is important. We suggest using tools and metrics from the complex network domain  to determine the best placement of the file replicas.

In particular, we have used a community detection algorithm to determine the regions where different replicas should be stored and a centrality index to chose the specific fog devices into a region. The selected fog devices should keep a trade-off between: storing the replicas closer enough to the sensors and the cloud provider, to reduce the latency and to increase the network as less as possible; but also evenly enough, to guarantee the file availability in the case that a region of the network (or any of the fog devices) gets down.

We define the placement algorithm by constraining it to three replicas of the files stored in three different devices.  Algorithm~\ref{placement} shows the pseudo-code of the process of placing the file replicas. It could be easily extended for a higher number of replicas or for considering file chunks stored in different devices.

Our algorithm first partitions the network into two separated regions (line~\ref{line:kernighanlin}). one region less than the replication factor. It is expected that the elements inside each region would be better communicated between them than between the elements in the other regions. Community detection algorithms are suitable solutions to find those partitions~\cite{FORTUNATO201075}. We consider that these regions or communities are analogous to the concept of the racks in the case of HDFS and data centers, but for the case of fog domains. We have named this policy as community-awareness (likewise to the rack-awareness name for HDFS).   Our algorithm would be extended to contexts with a higher replication factor by applying the partition algorithm sequentially to the resulting regions. For instance, in the case of a replication factor of five, the algorithm should need to partition the network into four regions in this first step. Consequently, two regions are obtained in a first iteration, and, by applying the algorithm again to each partition, we would finally obtain four regions.  

If only the devices and their connections are considered, the communities are static and equal for all the cases and conditions, and the same two regions/partitions are obtained for all the files. This is a drawback in the case, for example, that all the sensors of a given file are connected to gateways included in the same region. By this, the second replica would be placed too far from the sensors, increasing the network usage and latency. Consequently, we performed a different community detection process for each file with a complex weighted network also considering the distances to the sensors of a given file data type.

The weight of an edge for a given file is determined by the summation of the hop count between the edge and all the sensors of the file. The hop count is the number of intermediate devices, in the shortest path, through which data must pass between the sensor and the fog device. In the process of weighting complex networks of a given data type, it is necessary to: (i) identify all the subset of sensors which generate the given type of data for the file (line~\ref{line:getsensors}); (ii) sequentially calculate the hop count between all the edges and the sensors in the subset obtained in the previous step (line~\ref{line:weightgraph}); and finally, (iii) add the hop counts of an edge for each sensor (also line~\ref{line:weightgraph}). 

\begin{figure*}[!t]
	\centering
	\subfloat[Weighted graph for sensor A.]{\includegraphics[width=0.49\linewidth]{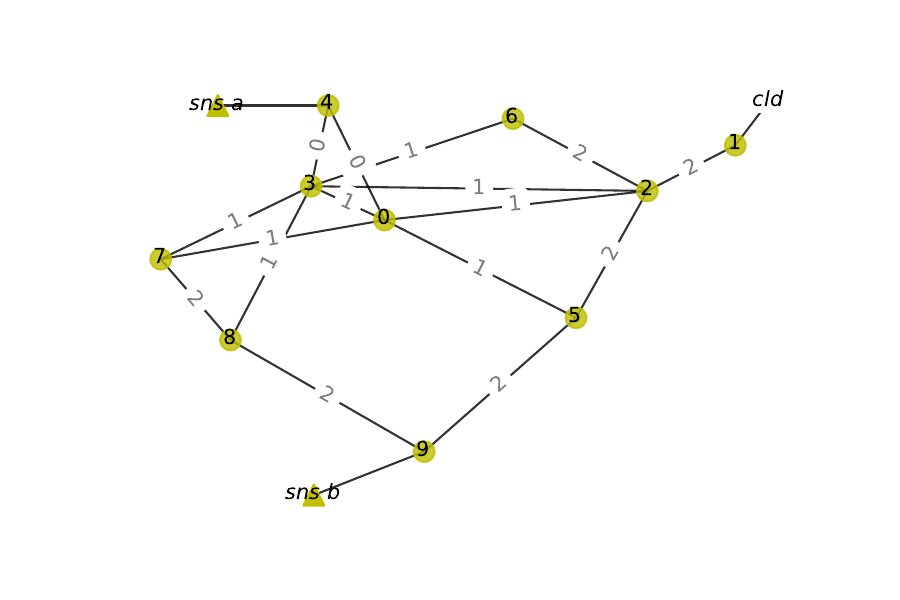}
		\label{fig:snsa}}
	\hfil
	\subfloat[Weighted graph for sensor B.]{\includegraphics[width=0.49\linewidth]{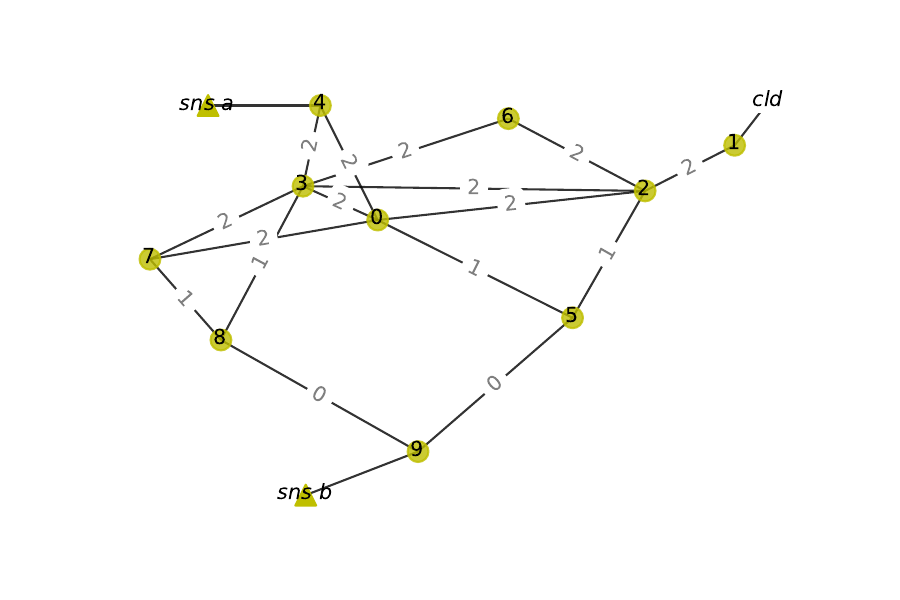}
		\label{fig:snsb}}
	\hfil	
	\subfloat[Weighted graph for both sensors]{\includegraphics[width=0.49\linewidth]{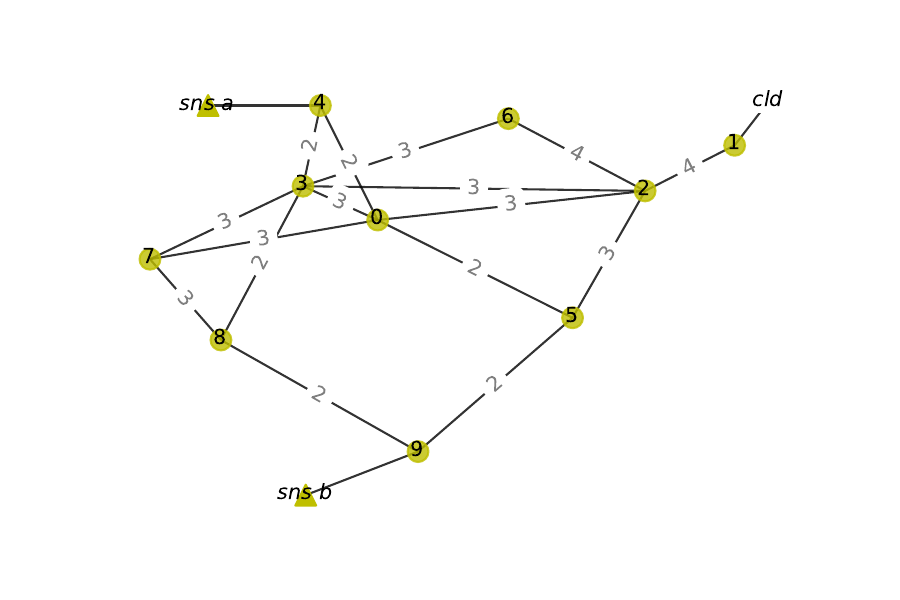}
		\label{fig:snstotal}}
	\hfil	
\subfloat[Result of the partition of the network]{\includegraphics[width=0.49\linewidth]{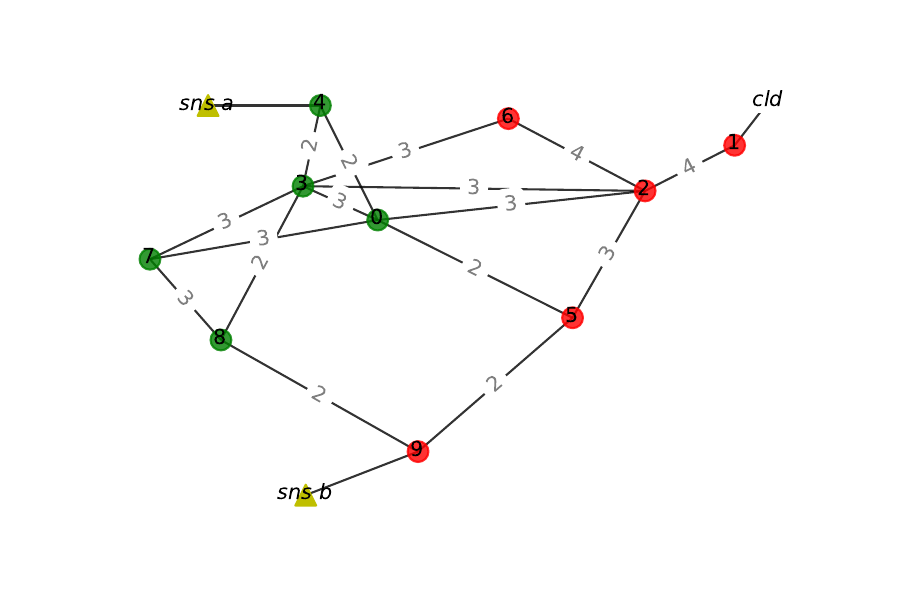}
	\label{fig:community}}	
	\hfil	
\subfloat[Placement of two replicas in the devices with the highest centrality values from each partition.]{\includegraphics[width=0.49\linewidth]{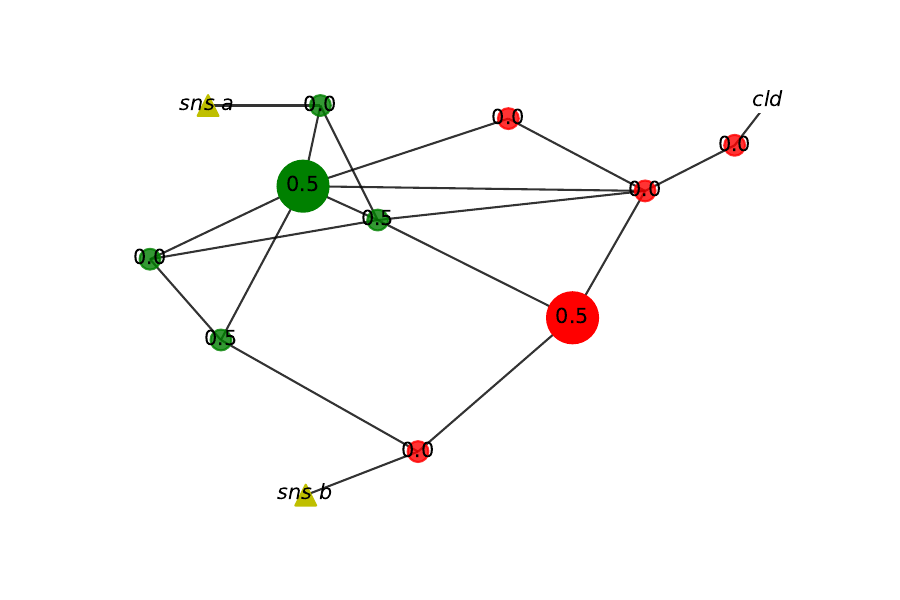}
	\label{fig:centralities}}
\hfil	
\subfloat[Placement of the third replica in the device with the highest centrality.]{\includegraphics[width=0.49\linewidth]{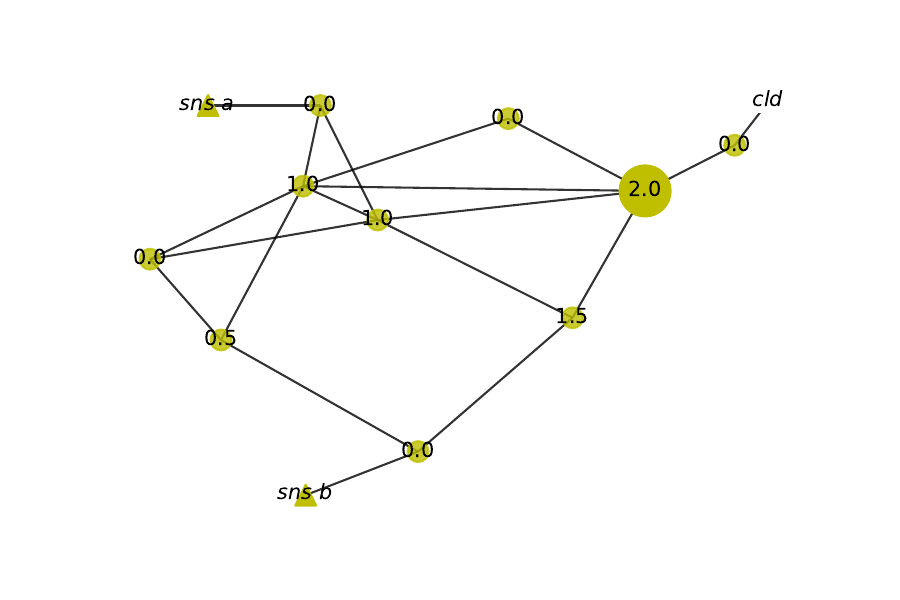}
	\label{fig:centralitiestotal}}	
	\caption{Example of the placement of the replicas of a file with two data sources.} \label{fig:allocationexample}
\end{figure*}

The weighting process is graphically represented in Figure~\ref{fig:allocationexample}. This example illustrates all the process for the placement of the replicas of a file with two data sources or sensors. Those sensors are attached to gateways 4 (sensor A) and 9 (sensor B). Figure~\ref{fig:snsa} shows the weights for the case of considering only sensor A and Figure~\ref{fig:snsb} for the case of sensor B. The final weighted graph is presented in Figure~\ref{fig:snstotal}, where the weights in the two previous cases have been added for each edge.

We partitioned the network into two blocks with the algorithm of Kernighan and Lin~\cite{6771089} (line~\ref{line:kernighanlin}). This algorithm partitions the nodes of a graph with costs on its edges into subsets of equal sizes so as to minimize the sum of the costs on all edges cut, i.e., to minimize the cost of the edges between the partitions. By this, the resulting two partitions minimize the number of network paths between devices which lie on different partitions, by also considering how far the devices are from the sensors of a given file. Consequently, the partitions are dynamically calculated for each of the files.

Figure~\ref{fig:community} shows the two partitions obtained for our example. The nodes are colored with green and red to represent the partition they belong to. Note that both partitions are of the same size and that one sensor is located in each of the partitions.

Once that two partitions of the network are determined for a given file, our algorithm places one replica of the file in each of the partitions. We determine the fog devices that would allocate the replica using the centrality value (line~\ref{line:centrality}). Centrality indices are indicators of the importance of the nodes in a graph~\cite{Newman}. We calculated the centrality values with the betweenness index~\cite{Koschuzki2005}. Betweenness centrality quantifies the number of times a node acts as a bridge along the shortest path between two other nodes. It measures the control of the node on the communication between the other two nodes. Although betweenness centrality showed pour results in our previous study for the placement of non-replicated files~\cite{8364053}, we have selected this index because it allows us to calculate the centrality for only a subset of target and source nodes. We need to determine the most important nodes not for all the network but only for the gateways with the sensors attached to. We used the betweenness centrality by using as sources and targets the subset of nodes that correspond to the gateways of the sensors. By this, the obtained centrality values are calculated in terms of the shortest paths between the sensors of a file.

Figure~\ref{fig:centralities} represents the nodes of the graph of our example with their betweenness centrality value. These values are obtained by considering only the shortest paths between the sensors (gateways 4 and 9) without the weights of the network. In this simple example, there are two shortest path between both gateways and the nodes in the shortest path has a centrality value of 0.5 because they are part of the half of the shortest paths.

The devices with the highest centrality value from each of the partition are selected as the candidate allocations (lines~\ref{line:highestcentrality1} and~\ref{line:highestcentrality2}). These devices would be finally selected if the free resources are higher than the resources that need to be provisioned to allocate the file replica. If not, the device with the next highest centrality would be evaluated for the placement. This process would be repeated until a device with enough resources is selected (from line~\ref{line:beginfreeresources} to line~\ref{line:endfreeresources}). 

In our simple example of Figure~\ref{fig:centralities}, there are several devices with the same centrality value of 0.5. In those cases, one random device can be selected. The example finally allocates the replicas in sensors 3 and 5 (represented as the nodes in the graph with the biggest size).

The placement of the two previous replicas is determined by the location of the data generators (sensors) in the network topology. But it is also important to reduce the distance from the replicas to the data consumers. As we explained in Section~\ref{system}, we suppose that the data are read by services or applications placed in the cloud provider. We named to this part of the algorithm as the cloud provider-aware placement policy, i.e., the location of the cloud provider is also considered for the replica placement. 

The placement of this third and last replica of each file is conducted to also reduce the distance between it and the cloud. Our algorithm calculates again the betweenness centrality (line~\ref{line:centralitywithcloud}) considering the shortest paths between the sensors but also, in this case, with the cloud provider (line~\ref{line:getsensorsandcloud}).   If we wanted to extend the algorithm for cases where the data consumer types are not only allocated in the cloud, we would calculate the value of the centrality index for the third replica considering the placement of the data consumer types instead of the placement of the cloud.  

The device with the highest centrality and enough resources is selected for the third replica placement. If the device already allocates any of the two first replicas, the device with the next highest centrality is selected (from line~\ref{line:beginfreeresources} to line~\ref{line:endfreeresources}). For instance, Figure~\ref{fig:centralitiestotal} shows that node 2 has the highest centrality for the subset of nodes 4 and 9 (the gateways with the sensors), and 1 (the cloud provider). Thus the third replica is placed in device 2.

 To sum up, our proposal is based on the idea that the placement of the two first replicas benefits the writing operations from the sensors. The placement of the third replica is addressed to find a trade-off between the writing operations from the users and the reading operations from the cloud.  

Finally, the heuristic process of determining the placement of the file replicas is performed sequentially and different file orderings could result in different solutions due to the limited capacities of the fog devices. We prioritized the replica placement by ordering the files by the total data generation rate of each file (line~\ref{line:fileorder}). Thus, the files with higher data generations are allocated before than the ones with smaller rates (line~\ref{line:fileorderfor}).

\begin{algorithm}[t!]
	\caption{Algorithm for replica placement in devices}\label{placement}
	\begin{algorithmic}[1]
		\Procedure{ReplicaPlacement}{G: network graph, FILES: set of files}
		\State \textit{/*replica placement ordered by the file's data generation rate*/}
		
		\State orderedFiles $\leftarrow$ order(FILES, by='$writeRate_f$')\label{line:fileorder}
		\For{$file_f$ \textbf{in} orderedFiles} \label{line:fileorderfor}
		\State subsetGateways $\leftarrow$ getSensorDataSources($file_f$)\ \ \ /*list of the sensors' gateways for file f*/ \label{line:getsensors}
		
		\State \textit{/*community-aware placement of the two first replicas*/}
		
		\State centrality $\leftarrow$ betweenness(DEV, sources=subsetGateways, targets=subsetGateways) \label{line:centrality}
		\State \textit{/*centralities considering the gateways with connected sensors*/}
		\State Gw $\leftarrow$ weightGraphWithHopCount(G) \label{line:weightgraph}
		\State rack0, rack1 $\leftarrow$ kernighanLinPartition(Gw) \label{line:kernighanlin}
		\State orderedRack0 $\leftarrow$ order(rack0, by='centrality') \label{line:highestcentrality1}
		\State orderedRack1 $\leftarrow$ order(rack1, by='centrality') \label{line:highestcentrality2}
		\For{$dev_i$ \textbf{in} orderedRack0} \label{line:beginfreeresources}
		\State \textit{/*search of the device with enough resources and highest centrality for the first partition*/}
			\If {freeResources($dev_i$) >= $dataReq_f$}
				\State firstAllocation $\leftarrow$ $dev_i$
				\State \textbf{break}
			\EndIf
		\EndFor
		\For{$dev_i$ \textbf{in} orderedRack1}
				\State \textit{/*search of the device with enough resources and highest centrality for the second partition*/}
			\If {freeResources($dev_i$) >= $dataReq_f$}
			\State secondAllocation $\leftarrow$ $dev_i$
			\State \textbf{break}
			\EndIf
			\EndFor \label{line:endfreeresources}

		\State \textit{/*cloud provider-aware placement of the last third replica*/}
		\State subsetDevices $\leftarrow$ subsetGateways$\cap$cloudProvider \label{line:getsensorsandcloud}
		\State centralityWithCloud $\leftarrow$ betweenness(G, sources=subsetDevices, targets=subsetDevices) \label{line:centralitywithcloud}
		\State \textit{/*centralities considering the cloud provider and the gateways with connected sensors*/}
		\State orderedDevices $\leftarrow$ order(DEV, by='centralityWithCloud')
		\For{$dev_i$ \textbf{in} orderedDevices} \label{line:beginfreeresourceswithcloud}
		\If {freeResources($dev_i$) >= $dataReq_f$ \textbf{and} $dev_i$ != firstAllocation \textbf{and} $dev_i$ != secondAllocation}
		\State thirdAllocation $\leftarrow$ $dev_i$
		\State \textbf{break}
		\EndIf
		\EndFor \label{line:endfreeresourceswithcloud}
		\State \textit{/*allocation of the replicas in the selected devices and update of the free resources*/}
		\State allocateReplica($file_f$, firstAllocation)
		\State allocateReplica($file_f$, secondAllocation)
		\State allocateReplica($file_f$, thirdAllocation)
		\EndFor
		\EndProcedure
	\end{algorithmic}
\end{algorithm}

\subsection{Single Replica Placement}
\label{singledata}

We proposed the use of the centrality indices for the placement of the files that store data generated in the sensors (data sources) of a fog architecture in a previous research work~\cite{8364053}. In this preliminary study, the availability of the data is not considered, and only one replica of each file is allocated in the devices. The objective of the algorithm is to store the single replica in the fog device with the closest and most evenly distance to the data sources (sensors), in order to optimize the network usage. We propose to: (i) model the network topology as a complex weighted network; (ii) calculate the centrality index of each node; (iii) select the node with the highest centrality index and enough free storage capacity to allocate the data generated for a given set of sensors of the same type. By this, we consider that the network usage is reduced and the data is stored as closer as possible of the data sources. As the capacities of the fog devices is limited, a priority rule need to be established and we decided to use the data generation rates: the higher the generation rate, the higher the priority.

The centrality of the devices is not only calculated with the topology of the network but also by weighting the edges. We have weighted the edges with the summation of the hop count to all the sensors of the given data type, in the same way that we explained in Section~\ref{replicadata}. By this, we prioritize the closest nodes and penalize the most distant ones. The hop count is the number of intermediate devices through which data must pass between the sensor and the fog device.

This preliminary approach was validated by simulating several experimental scenarios. The experiments were used to compare three centrality indices (Eigenvector, Betweenness and Current Flow Betweenness) in four network topologies (Barabasi-Albert, Grid, Random Euclidean and Lobster). The results were compared with control case of storing all the data in the cloud provider. The experiment were also scaled by considering different number of sensors.

In general terms, the eigenvector centrality showed better results than the other centrality indices. The betweenness centrality was the worst policy, even obtaining worst results than the placement in the cloud for many of the experiments. The current flow index obtained a similar result than the cloud placement, since it  better for some experiments and worst for others. 

Consequently, we select the eigenvector centrality index for the single replica placement. On contrary, we select the betweenness centrality for the replica-aware data placement although its pour results because, as we explained in Section~\ref{replicadata}, this index allows to select source and target nodes.

\section{Experimental Evaluation}
\label{experimental}

The evaluation of our proposal is based on measuring our optimization objectives ($filAvl_{read}$,  $filAvl_{write}$,  $accLat_{read}$, $accLat^{max}_{write}$, $accLat^{min}_{write}$,  $msgTrns_{write}$, and $msgTrns_{read}$) in a reference simulation scenario, whose size has been modified in terms of the number of files and number of devices. The experiments were executed using the YAFS simulator that we had previously developed for other research works. This simulator is able to include graph-based network topologies and pluggable fog service placement policies, apart from other features that, to the best of our knowledge, are not provided by other fog simulators, such as node failures, or dynamic service placement and routing. The simulator is open source and it can be downloaded from its code repository~\cite{yafs}.

We also implemented two additional placement policies in the experiments to compare our approach with another policy that considers replicas and with a policy that only stores one replica. Our policy is labeled as \textit{replica-aware}. The other policy with replication of the files is based on FogStore framework proposed by Mayer et al.~\cite{DBLP:conf/fwc/MayerGSR17}, and it is labeled as \textit{fogstore}. Finally, the single replica policy is the one in our preliminary study~\cite{8364053}, and it is labeled as \textit{single-file}.

In the case of FogStore, as we mentioned in Section~\ref{related}, it only considers one data source and it is human-assisted  which requires that the partition of the network has to be provided to the algorithm. We have solved these issues by first randomly selecting just one of the data source and by using a community detection algorithm for the partition into regions.  In the case of the \textit{single-file} policy, as we explained in Section~\ref{singledata}, we have selected the eigenvector centrality index since it showed the best results our previous study.

A reference experiment configuration has been defined and modified in terms of number of devices and number of files in the system to study the behavior of our placement policy. We considered a total number of 20 experiment sizes:   10 experiments with 100 files ranging the number of fog devices from 100 to 300 fog devices in steps of 20; and 10 more experiments with 200 fog devices and the number of files ranging from 100 to 200 in steps of 10.  

This design choice is based on the idea of having experiments ranging from a case where the total storage resource usage of the fog devices is very low to a case where almost the total storage capacity of the fog devices is used. The column \textit{Storage usage ratio} of Table~\ref{tab:improvementlatencysensorreplica} shows the ratio between the total storage capacity required by the files and the total storage capacity offered by the fog devices. The experiments range from storage usages around 30\% to cases around 85\%.

The summary of these and other characterization parameters of the reference experiment are presented in Table~\ref{tab:experimentparameters}. All the parameters have been generated with a uniform distribution to evenly cover all the values in the range between the minimum and maximum values.

We have modeled the network as a Barabasi-Albert topology since this is a common network topology~\cite{8368525}.  The $gtwPercentage$ percent of the nodes with the smallest betweenness centrality were selected to be the gateways (fog devices with connected sensors). Once the gateways were selected, the sensors for each data type (file) were connected to them. The sensors were created following a popularity uniform distribution whose parameter was randomly selected for each file with a value between 0 and $snsPopularity$. Finally, a new node was introduced to act as the cloud provider and connected to all the fog devices with the highest betweenness centrality.

\begin{table}
	\centering
	\caption{Values of the parameters for the characterization of the experiments}\label{tab:experimentparameters}
	

		\rowcolors{2}{gray!20}{white}
		\begin{tabular}{llr}
\rowcolor{gray}

\textbf{Parameter} &  & \textbf{Min. and max. values} \\
\multicolumn{1}{l}{\textbf{Network}} &&\\
Number of fog devices & |DEV| &  100--300 \\
Propagation time (ms) & $netPrp_j$& 1--5\\
Bandwidth (bytes/ms)  & $netBdw_j$& 50000--75000\\
\multicolumn{1}{l}{\textbf{Fog device}} &&\\
Resources (storage space) & $datCap_u$& 10--25\\
Percentage of gateways & $gtwPercentage$& 10\%  \\
\multicolumn{1}{l}{\textbf{Files}} &&\\
Number of files & |FILE| &  100--200 \\
Replication factor &&3 \\
Resources (storage size)&$datReq_f$& 1--6\\
Read packet size (bytes)&$readPacketSize_f$& 1500000--4500000\\
Write packet size (bytes)&$writePacketSize_f$& 1500000--4500000\\
\multicolumn{1}{l}{\textbf{IoT device}} &&\\
Write request rate (1/ms)&$writeRate_f$& 1/1000--1/200\\
File popularity& $snsPopularity$ &   15\%  \\
\multicolumn{1}{l}{\textbf{Cloud provider}} &&\\
Read request rate (1/ms)&$readRate_f$& 1/6000--1/1200\\

\bottomrule

\end{tabular}%

\end{table}

\section{Results and Analysis}
\label{resultanalysis}

The results of the simulation and their analysis are presented in terms of the three optimization objectives: the network latency, the network usage and the file availability. The simulations were repeated 10 times to avoid deviations in the averages. The standard deviations were calculated and it resulted in low values that guaranteed the significance of the results. The title of each plot includes the maximum standard deviation of the metrics that it represents.

The following subsections analyze and compare the results of our proposed policy for the placement of replicas of a file (\textit{replica-aware}), compared to other two placement policies: \textit{single-file} which only stores one replica of the files, and \textit{fogstore} based on the work of Mayer et al.~\cite{DBLP:conf/fwc/MayerGSR17}.

Additionally to the graphical representation of the results, we have also included a summary of the results in Table~\ref{tab:improvementlatencysensorreplica}, where the improvement ratios of our policy in regard with the results of the FogStore are compared. The improvement ratio is calculated as the division between our policy metric value and the FogStore one ($\frac{replica-aware\ metric\ value}{fogstore\ metric\ value}$). This ratio indicates the percentage of improvement of our policy in regard to FogStore. For example, an improvement ratio of 1.2 indicates that our policy is a 20\% better than FogStore.

\subsection{Network Latency}
\label{sect:networklatency}

\begin{figure*}[!t]
	\centering
	\subfloat[Experiments with a fixed number of 200 fog devices and a variable number of files.]{\includegraphics[width=0.98\linewidth]{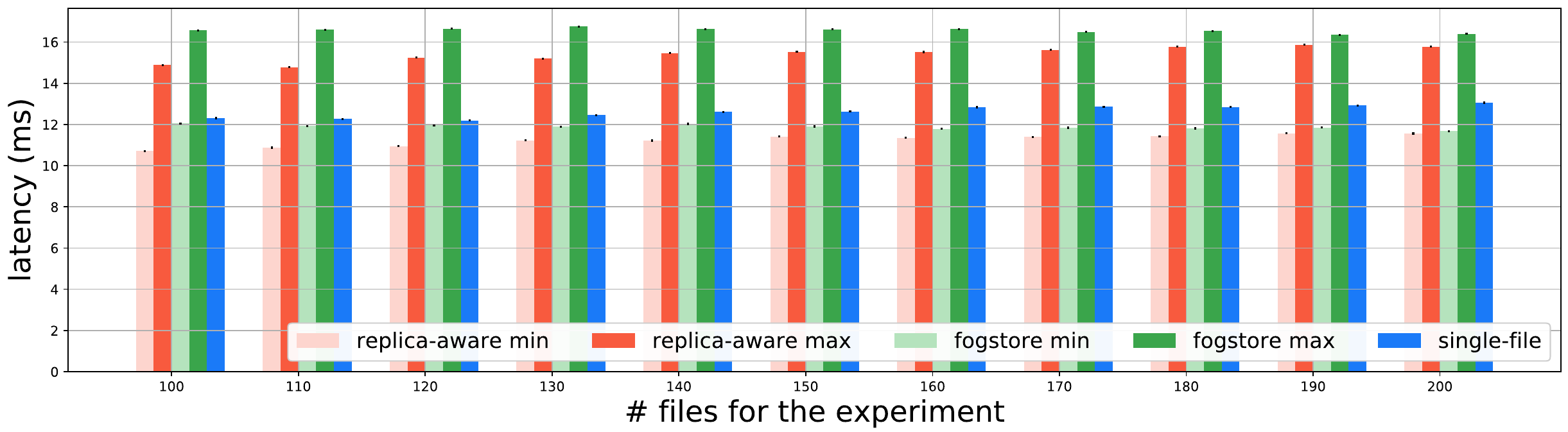}
		\label{fig:latencysensorreplica50devices}}
	\hfil
	\subfloat[Experiments with a fixed number of 100 files and a variable number of fog devices.]{\includegraphics[width=0.98\linewidth]{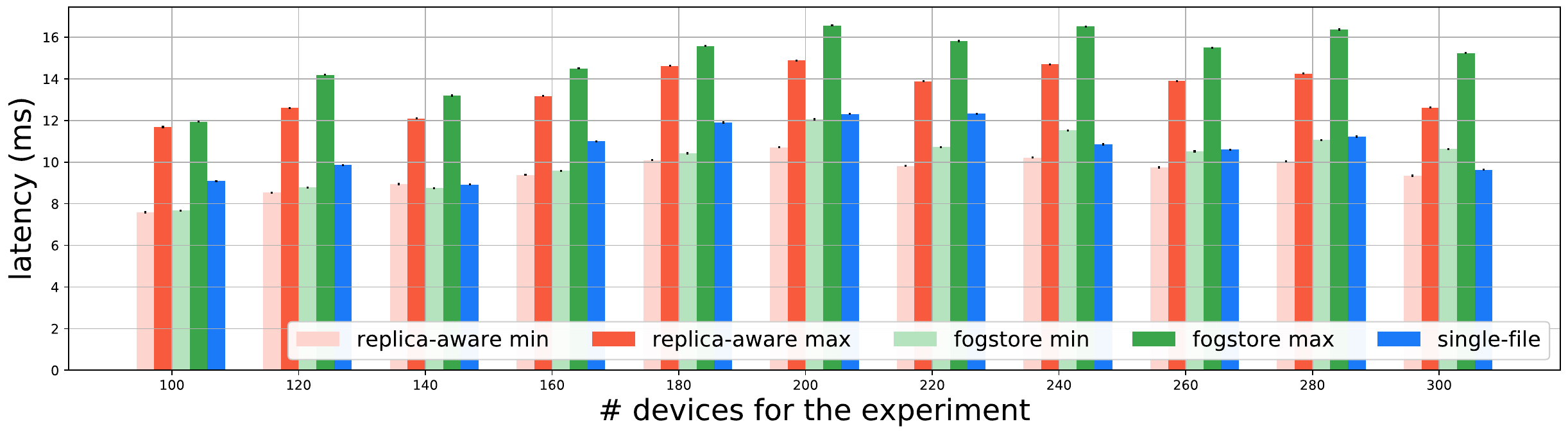}
		\label{fig:latencysensorreplica100files}}
	\hfil	
	\caption{Results for the total latency for writing operations between the sensors and the file replicas.} \label{fig:latencysensorreplica}
\end{figure*}

\begin{figure*}[!t]
	\centering
	\subfloat[Experiments with a fixed number of 200 fog devices and a variable number of files.]{\includegraphics[width=0.98\linewidth]{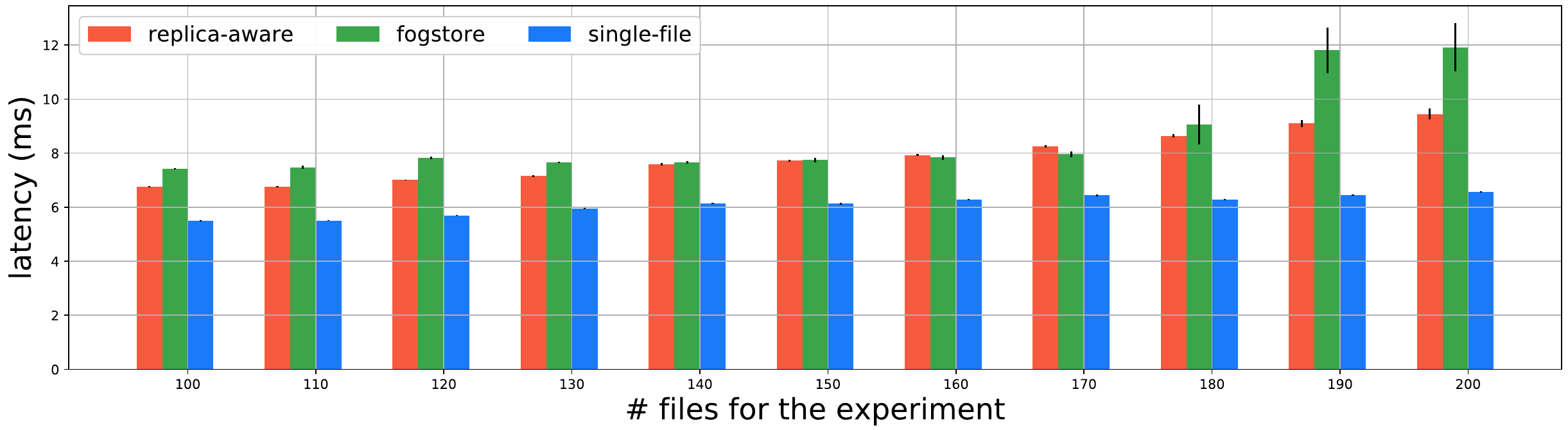}
		\label{fig:latencyreplicacloud50devices}}
	\hfil
	\subfloat[Experiments with a fixed number of 100 files and a variable number of fog devices.]{\includegraphics[width=0.98\linewidth]{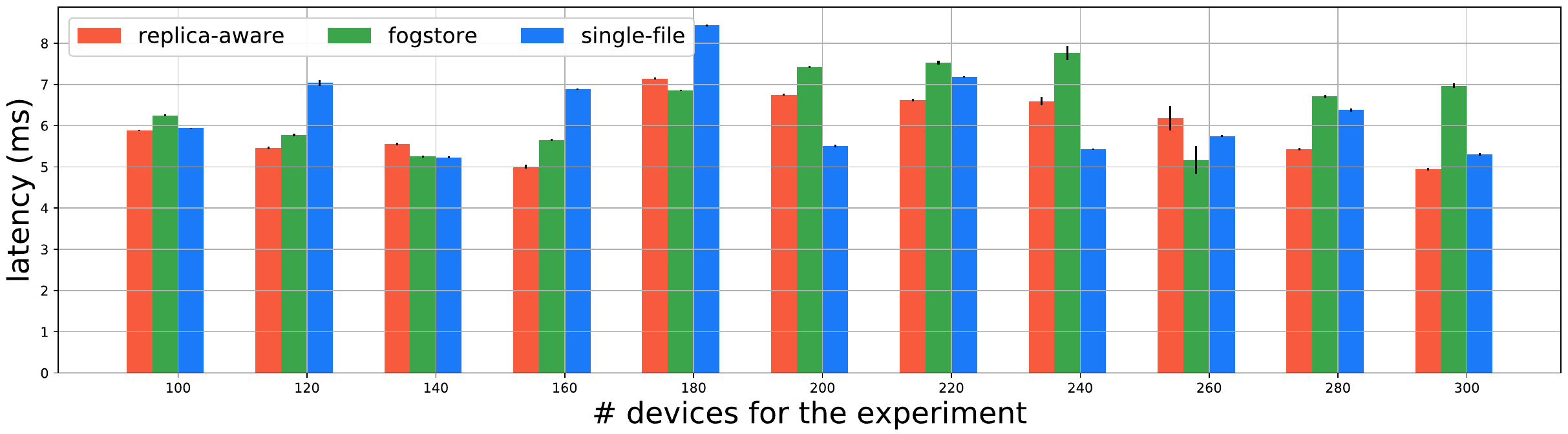}
		\label{fig:latencyreplicacloud100files}}
	\hfil	
	\caption{Results for the total latency for reading operations of the cloud provider in the file replicas.} \label{fig:latencyreplicacloud}
\end{figure*}

The results for network latency are presented separately for the latency between the sensors and the file replicas (writing operations), in Figure~\ref{fig:latencysensorreplica}, and between the replicas and the cloud provider (reading operations), in Figure~\ref{fig:latencyreplicacloud}.

It is important to remember and remark that writing operations, for the case of the replica-aware policy, are performed on three different locations, and we simulated that those operations were performed in parallel. Under these conditions, we have considered that the best way to represent the latency is by analyzing the cases for the closest replica (labeled as \textit{replica-aware/fogstore min} in Figure~\ref{fig:latencysensorreplica}) and the furthest one (labeled as \textit{replica-aware/fogstore max}). Thus, both latencies give an idea of the range of the times for writing operations. On the contrary, in the case of the single-file policy, the file is stored only in one device, and it is not necessary to differentiate between those both latency times.

As it was expected, the latency times for the single-file are measured in between the minimum and maximum values of the   other two replicas with replicas. It is important to remember that the replica-aware spreads the file replicas across the architecture to improve the file availability and, consequently, it is quite clear that the latencies of the furthest replicas are increased with regard to a policy which stores the files in the most central nodes, as the single-file policy does.    In regard to the FogStore, our policy shows smaller maximum and minimum latencies for writing operations, and it is, in average, 5.7\% and 6.4\% better than FogStore for the minimum and maximum latencies respectively (Table~\ref{tab:improvementlatencysensorreplica}).

In the second case, Figure~\ref{fig:latencyreplicacloud}, we simulated that the reading operations were only performed on one of the replicas of the file, in particularly, which was closest to the cloud provider. Thus, only the time of the closest replica is represented in the figure for the case of the replica-aware policy. All the experiments show that our proposed policy obtained shorter latency times for reading than the single-file policy, except for one experiment (200 devices and 100 files).   In regard to the FogStore, our policy again shows shorter reading latencies, except for three cases (the three cases with values smaller than 1.0 in the column \textit{IR reading latency} in Table~\ref{tab:improvementlatencysensorreplica}). In average, our policy is around 12\% better than the FogStore for reading operations.

If we analyze the evolution of the improvement rations in regard to the percentage of the storage required for the files (column \textit{Storage usage ratio} in Table~\ref{tab:improvementlatencysensorreplica}), a certain correlation is observer between them. The improvement ratio of policy is higher (higher benefits) when the percentage of storage usage is lower, and the benefits are reduced as the storage usage is increased. This is because as the storage usage is increase, the number of devices that are fully used is also increased, and the files need to be stored in non-optimal places due to the resource usage constraints. Consequently, the differences between policies are softened.

In general terms, we can conclude that our policy is able to minimize the latency times for reading operations, and in the case of the writing ones, only the latencies of the furthest replicas are slightly longer than the latencies for the control experiment, the single-file policy.

\subsection{Network Usage}

In the case of the network usage, the results are also presented and analyzed separately for the case of the writing and reading operations.

\begin{figure*}[t]
	\centering
	\subfloat[Experiments with a fixed number of 200 fog devices and a variable number of files.]{\includegraphics[width=0.98\linewidth]{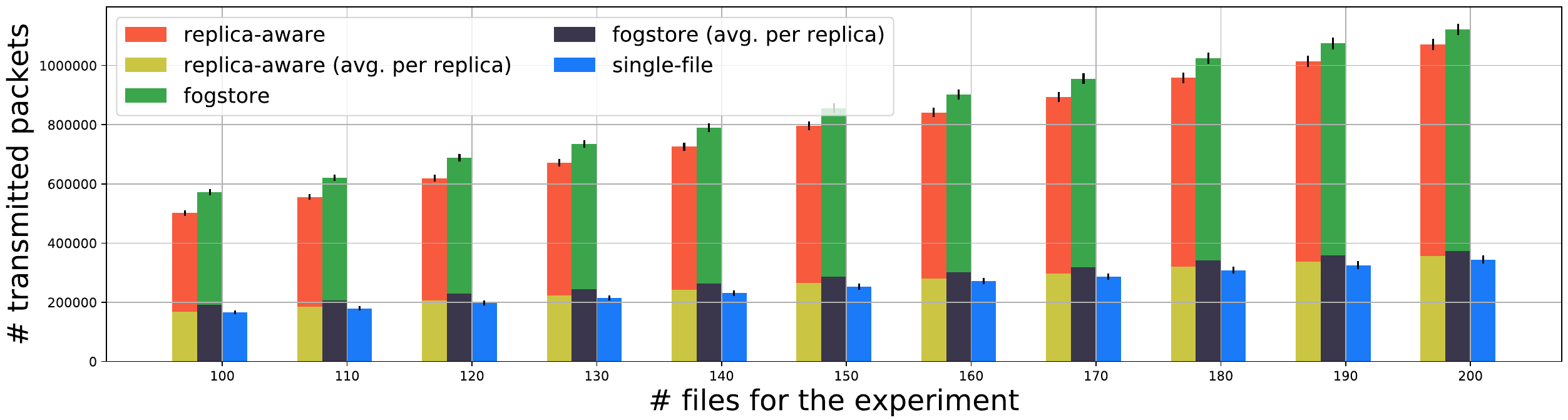}
		\label{fig:packetssensorreplica50devices}}
	\hfil
	\subfloat[Experiments with a fixed number of 100 files and a variable number of fog devices.]{\includegraphics[width=0.98\linewidth]{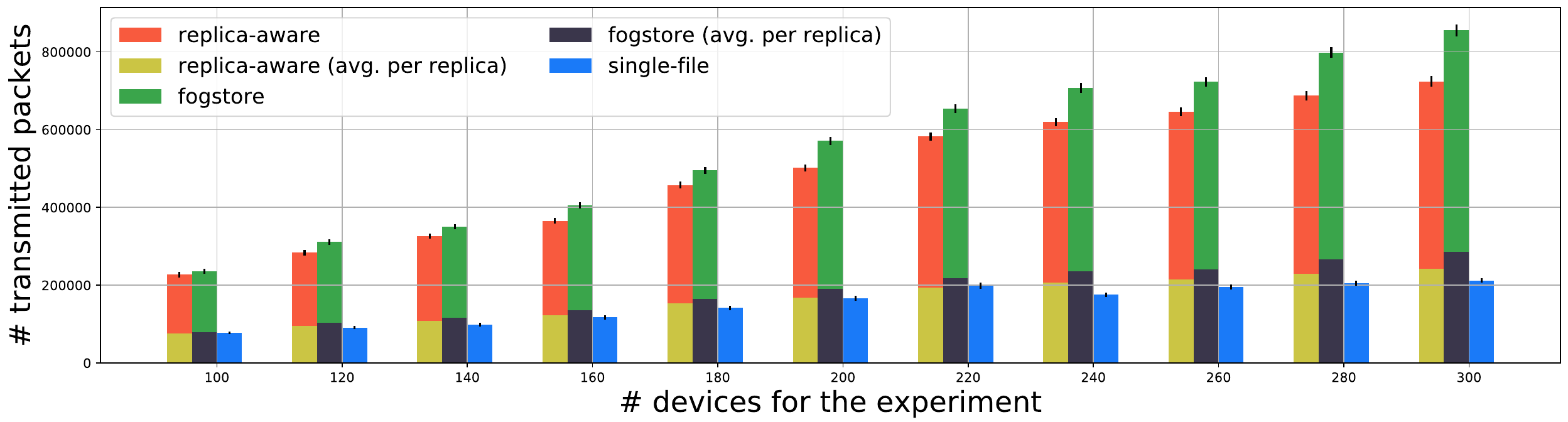}
		\label{fig:packetssensorreplica100files}}
	\hfil	
	\caption{Results for the number of transmitted packets for writing operations between the sensors and the file replicas.} \label{fig:packetssensorreplica}
\end{figure*}

In the case of the writing operations, Figure~\ref{fig:packetssensorreplica} measures and presents the total number of packets that are transmitted from the sensors to the file storages. The writing operations are performed in parallel in the case of the replica-aware policy and, consequently, the same packet is transmitted three times from the sensors to the replicas. Thus, it would be reasonable to obtain a number of packet three times bigger to the single-file policy for the policies that store replicas. For a deeper comparison between them, we have also represented the third of the total number of transmitted packets with both the \textit{replica-aware} and \textit{fogstore} policies. This is represented highlighted (in olive and dark colors) in front of the bars that correspond to the total number (red and green colors).

Although some slightly differences, it is generally observed that the number of packets using our policy is three times bigger than the single-file one, as it was expected. This could be optimized by considering to transmit the data from the sensors to the replicas sequentially, by sending data between replicas, instead of between sensors and replicas. By this, the network usage would be probably reduced, but the latency would be damaged because the three writing operations would be performed sequentially, from the sensors to the first replica, from the first replica to the second one, and, finally, from the second to the third one.

If we compare the results of our policy with the ones obtained with FogStore, it is observed that our policy reduce the number of transmitted packets in all the experiments for writing operations. And the average improvement ratio is 1.098, i.e., our policy is 9.8\% better than FogStore. It is also observed that the improvement ration is inversely proportional to the resource usages.

\begin{figure*}[t]
	\centering
	\subfloat[Experiments with a fixed number of 200 fog devices and a variable number of files.]{\includegraphics[width=0.98\linewidth]{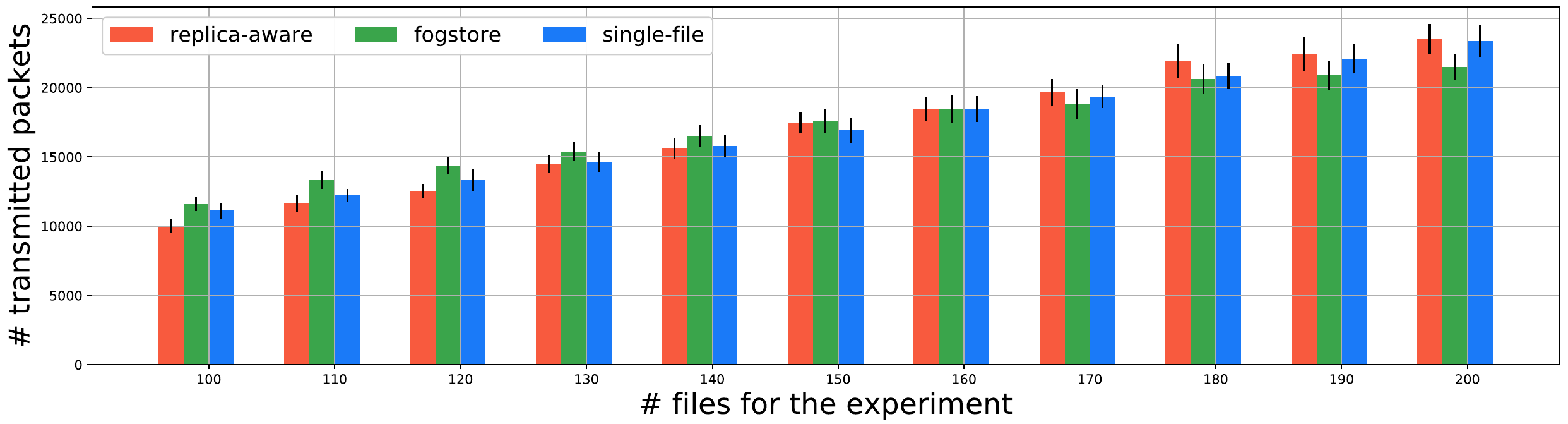}
		\label{fig:packetsreplicacloud50devices}}
	\hfil
	\subfloat[Experiments with a fixed number of 100 files and a variable number of fog devices.]{\includegraphics[width=0.98\linewidth]{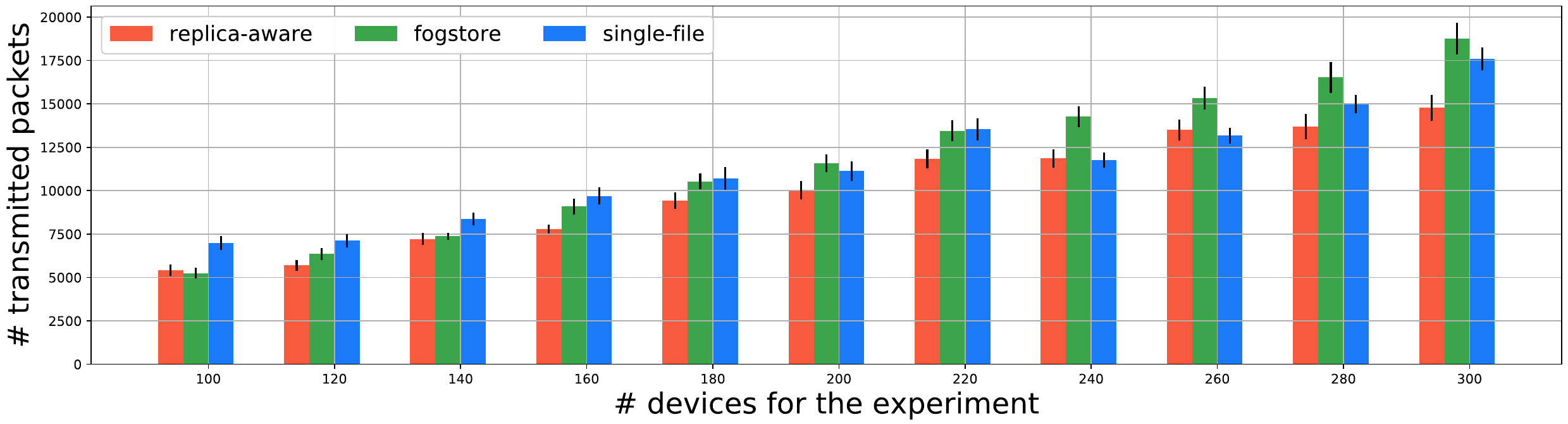}
		\label{fig:packetsreplicacloud100files}}
	\hfil	
	\caption{Results for the number of transmitted packets for reading operations of the cloud provider in the file replicas.} \label{fig:packetsreplicacloud}
\end{figure*}

As in the case of the network latency (Section~\ref{sect:networklatency}), we consider that the cloud only reads its closest replica. Thus, the simulation only transmits data (i.e., packets) from the closest replica to the cloud. Figure~\ref{fig:packetsreplicacloud} represents the total number of packets transmitted for reading operations in all the simulation.   Our policy is better than the \textit{single-file} and \textit{fogstore} results in almost all the cases, mainly in the experiments that vary the number of files (Figure~\ref{fig:packetsreplicacloud50devices}). More precisely, our policy is 12.5\% better than FogStore. Once again, it is observed a inverse correlation between the benefits of our policy and the storage usage.

In general terms, we can conclude that our policy increases the network usage because the data generated in the sensors need to be stored in three replicas allocated in three spread devices. But, although that, this increase is smaller than three times the number of packets of the single-file policy.   Additionally, our policy increases the network usage less than the other policy with file replicas, FogStore.  

\begin{table}
	\centering
	\caption{Improvement ratios (IR) of our \textit{replica-aware} placement policy in regard with the \textit{fogstore} for the latency and network usage metrics, and resource usages for each experiment. Resource usage ratio for each experiment.}\label{tab:improvementlatencysensorreplica}
	
	
	\rowcolors{2}{white}{gray!20}
	\begin{tabular}{rrrrrrrr}
		\rowcolor{gray}
		
		\textbf{Files} & \textbf{Nodes} & \textbf{IR writing} & \textbf{IR reading} & \textbf{IR reading} & \textbf{IR min. writing} & \textbf{IR max. writing} & \textbf{Storage}\\
		\rowcolor{gray}
		\textbf{} & \textbf{} & \textbf{messages} & \textbf{messages} & \textbf{latency} & \textbf{latency} & \textbf{latency} & \textbf{usage ratio}\\
		
100 &   100 &           1.047 &           0.996 &           2.767 &   0.972 &   1.016 & 0.848\\
100 &   120 &           1.092 &           1.017 &           1.158 &   0.973 &   1.093 & 0.685\\
100 &   140 &           1.067 &           0.924 &           1.042 &   0.912 &   1.091 & 0.586\\
100 &   160 &           1.101 &           1.167 &           0.797 &   1.028 &   1.052 & 0.518\\
100 &   180 &           1.082 &           1.083 &           1.051 &   0.969 &   1.039 & 0.471\\
100 &   200 &           1.134 &           1.200 &           1.042 &   1.227 &   1.081 & 0.423\\
100 &   220 &           1.124 &           1.266 &           1.111 &   1.164 &   1.103 & 0.381\\
100 &   240 &           1.148 &           1.150 &           0.994 &   1.188 &   1.109 & 0.351\\
100 &   260 &           1.129 &           1.421 &           1.220 &   0.982 &   1.090 & 0.323\\
100 &   280 &           1.160 &           1.236 &           1.084 &   1.187 &   1.149 & 0.299\\
100 &   300 &           1.182 &           1.204 &           1.014 &   1.134 &   1.281 & 0.280\\
		
		\midrule		
		
   100 &   200 &           1.134 &           1.200 &           1.042 &   1.227 &   1.081 & 0.423 \\
   110 &   200 &           1.115 &           1.207 &           1.037 &   1.123 &   1.058 & 0.467\\
   120 &   200 &           1.114 &           1.175 &           1.054 &   1.109 &   1.043 & 0.515\\
   130&   200 &           1.092 &           1.156 &           1.056 &   1.052 &   1.034 & 0.556\\
   140 &   200 &           1.091 &           1.126 &           1.040 &   1.056 &   1.019 &0.597 \\
   150 &   200 &           1.072 &           1.108 &           1.021 &   1.045 &   1.012 & 0.626\\
   160 &   200 &           1.064 &           1.087 &           1.031 &   1.002 &   1.019 & 0.662\\
   170 &   200 &           1.062 &           1.056 &           1.018 &   1.031 &   1.011 & 0.702\\
   180 &   200 &           1.060 &           1.038 &           1.019 &   0.999 &   1.009 & 0.745\\
   190 &   200 &           1.054 &           0.993 &           0.994 &   0.945 &   1.013 & 0.794\\
   200 &   200 &           1.046 &           0.942 &           1.008 &   0.933 &   1.016 & 0.837\\

		\bottomrule
		
		   \multicolumn{2}{c}{average}&           1.098 &           1.125 &           1.118 &   1.057 &   1.064 &  \\

	\end{tabular}%

\end{table}

\subsection{File Availability}

Once again, file availability is analyzed in terms of the writing operations that are carried out by the sensors in the three replicas of the files and in terms of the reading operations that the cloud provider performs over those replicas. As we mention in Section~\ref{system}, we consider that a given file is available if: (a) the cloud is able to reach at least one of its three replicas, in the case of the readings;  and (b) if all the sensors of the data source type of the given file are able to reach at least one replica.

For the evaluation of the file availability, we have measured the metrics $filAvl_{read}$ and $filAvl_{write}$ when a set of fog devices are down due to failures. Concretely, we have generated random failures in the 10\% of the fog devices and, consequently, some replicas and some network paths became unavailable for the sensors or the cloud provider. 

\begin{figure*}[!t]
	\centering
	\subfloat[Experiments with a fixed number of 200 fog devices and a variable number of files.]{\includegraphics[width=0.98\linewidth]{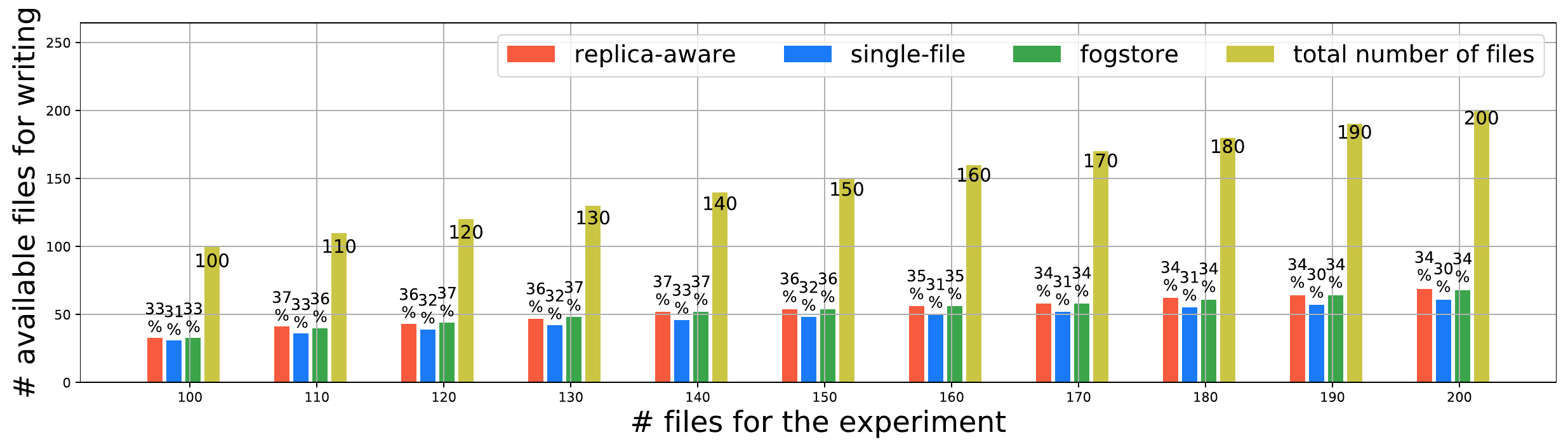}
		\label{fig:sensorwrite50devices}}
	\hfil
	\subfloat[Experiments with a fixed number of 100 files and a variable number of fog devices.]{\includegraphics[width=0.98\linewidth]{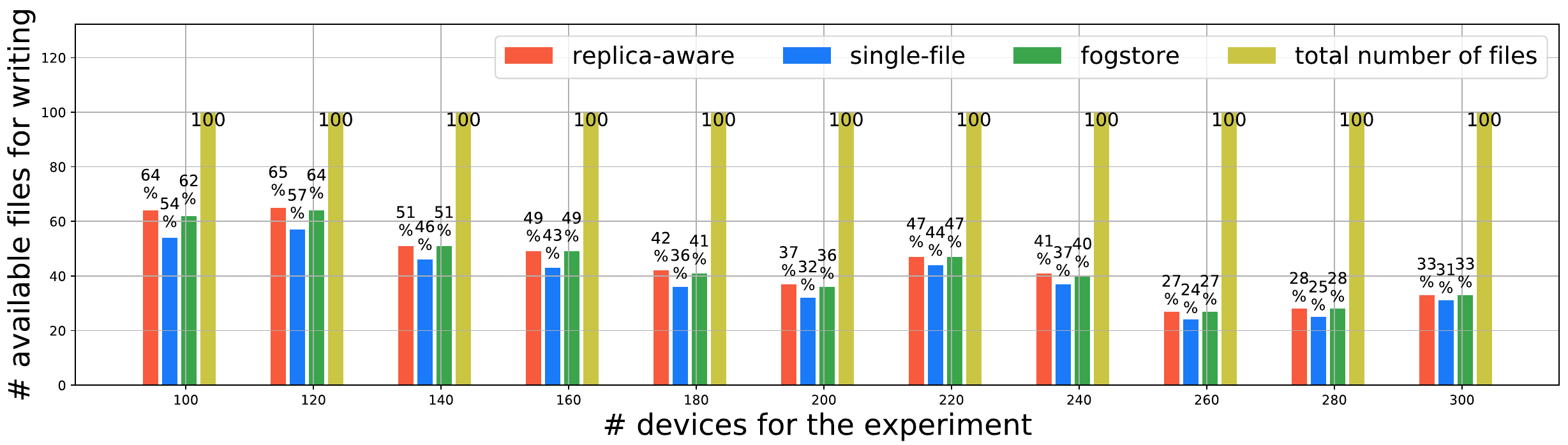}
		\label{fig:sensorwrite100files}}
	\hfil	
	\caption{Results for the file availability for writing operations from the sensors ($filAvl_{read}$) with failures in the 10\% of the devices.} \label{fig:sensorwrite}
\end{figure*}

Figure~\ref{fig:sensorwrite}  represents the number of files that are available for writing operations of the sensors for all the experiment sizes we considered. Additionally, the percentage of available files in the system is written in the top of the bars.

  It is clear that replica-based policies (our proposal and FogStore) show a higher number of available files after the 10\% of the devices failed. The differences with the single-policy are quite important, mainly for the reading operations. If we compare the two replica-based policies, the availability is very similar, without a clear best alternative. Patterns in regard to the experiment sizes or the storage usage are not detected in the case of the availability.

\begin{figure*}[!t]
	\centering
	\subfloat[Experiments with a fixed number of 200 fog devices and a variable number of files.]{\includegraphics[width=0.98\linewidth]{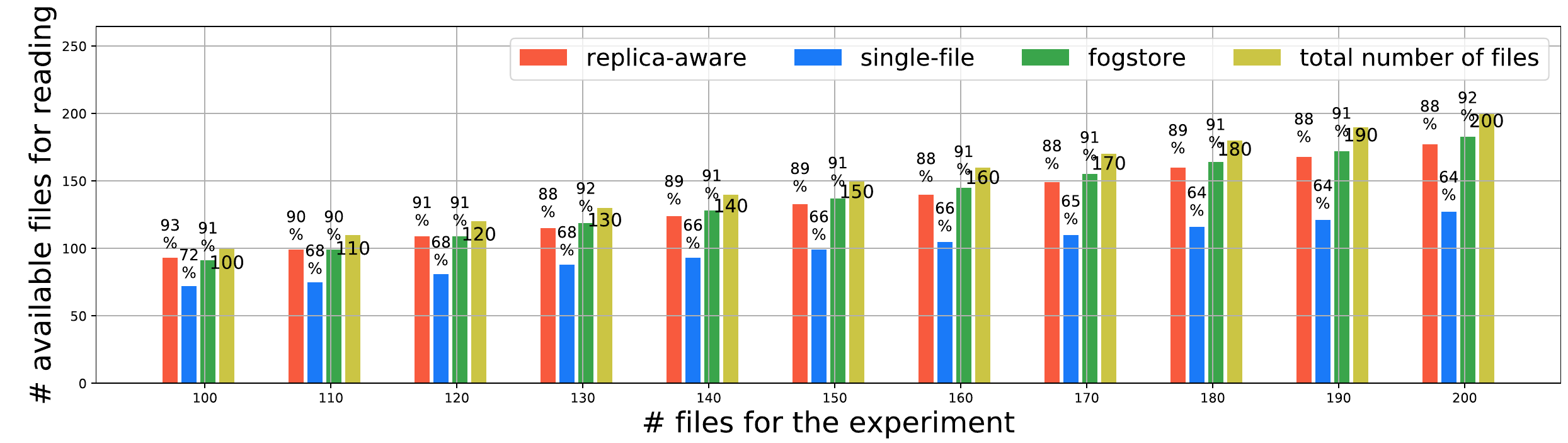}
		\label{fig:cloudread50devices}}
	\hfil
	\subfloat[Experiments with a fixed number of 100 files and a variable number of fog devices.]{\includegraphics[width=0.98\linewidth]{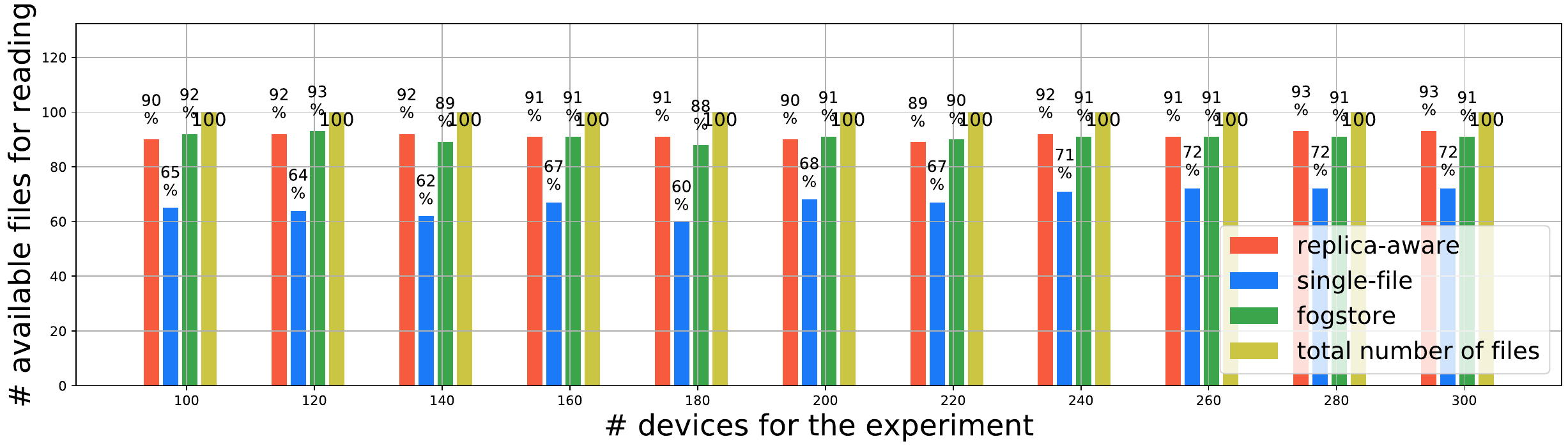}
		\label{fig:cloudread100files}}
	\hfil	
	\caption{Results for the file availability for reading operations from the cloud provider ($filAvl_{read}$) with failures in the 10\% of the devices.} \label{fig:cloudread}
\end{figure*}

Figure~\ref{fig:cloudread} represents the results for the second case, the reading operations from the cloud provider. The percentage of available files is also shown in the figures. In this case, it is observed that our policy is able to keep almost all the files available, with availabilities higher than 90\% for all the experiments. On the contrary, the single-file policy shows worse results, with results even smaller than the 60\% of availability.

To sum up, we can conclude that both policies show better availability for reading operations than for writing ones. For this last case, our policy is nearly able to keep all the files available for the cloud provider even after failures in the 10\% of the devices. In the case of the writing operations, our policy is again better than the single-file one, but with worse availability percentages. This is also explained by the strictness of our definition of the writing availability: if just one sensor is not able to reach a file, the file is computed as unavailable, even when other sensors would be able to write in this file.

\subsection{Placement Execution Time}

The applicability of the solution in large-sized architecture is also evaluated through the execution time of the placement algorithm. In a real scenario, the placement of the files is performed sequentially, i.e., the placement of each file is determined when it is created or when the number or position of the data sources (sensors) change. Consequently, the number of files does not influence in the scalability of the solution. On the contrary, the size of the network (number of fog devices) directly affects the execution time, since our algorithm requires the calculation of metrics of the topological structure of the network.

\begin{figure*}[!t]
	\centering
	\includegraphics[width=0.98\linewidth]{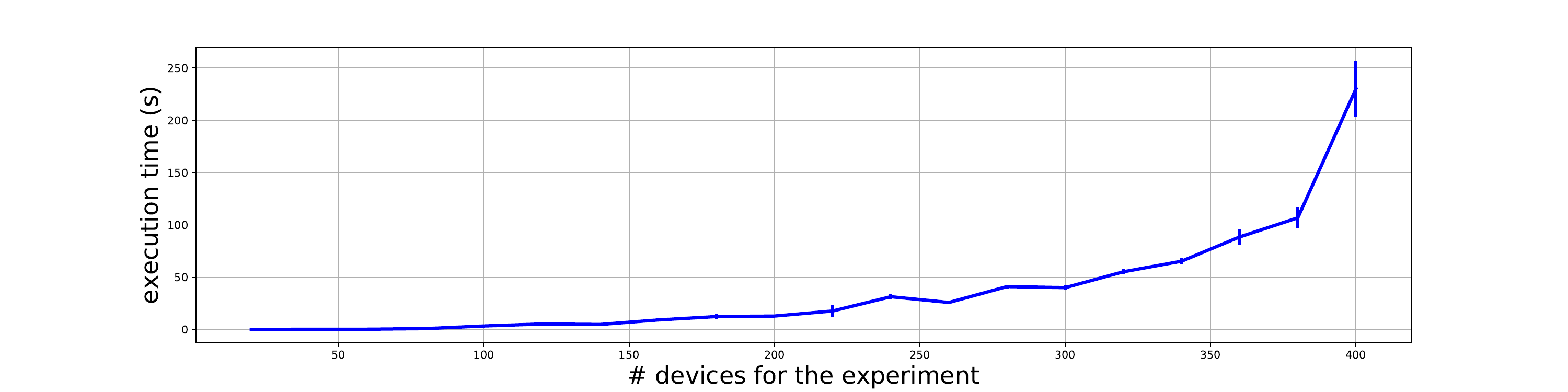}	
	\caption{Results for the file availability for writing operations from the sensors ($filAvl_{read}$) with failures in the 10\% of the devices.} \label{fig:executiontime}
\end{figure*}

Figure~\ref{fig:executiontime} shows the averages and standard deviation of the execution times to determine the placement of one file considering several network sizes. The algorithm has been implemented in Python 2.7.6 and it was executed on a computer running MacOS Sierra with an Intel(R) Core(TM) i7 processor operating at 3.10 GHz with 16 GB of RAM.

We can see that the execution time is worst than lineal. This is because the computational complexity of the topological metric are: $O(pn^2 \log n)$ for the Kernighan-Lin algorithm algorithm, where $n$ is the number of nodes in the graph (devices) and $p$ the number of iterations of the algorithm~\cite{Dutt:1993:NFK:259794.259855}; and $O(nm + n^2\log n)$ where $n$ is the number of nodes and $m$ the number of edges in the graph~\cite{doi:10.1080/0022250X.2001.9990249}. 

The applicability of the algorithm in high scale scenarios would be determined by the ratio of changes in the distribution of the data sources. High execution times could be allowed in static scenarios without changes in the sensors. But, as the number of changes in the sensor is increased, the placement could result unaffordable. For these high dynamic cases, the creation of data placement sub-problems for different geographical zones of the fog architecture could reduce the execution times. This would be specially interesting for cases in which the same data types are located together in the same region. In fact, several studies have shown that data in IoT domains is commonly generated locally~\cite{vanderZee2014}.

\section{Conclusion}
\label{conclusion}

We have presented a replica management policy to store the data generated in the sensors of an IoT system by using files placed in the fog devices. Our proposal is based on a community-awareness policy that balances the proximity of the file replicas to the data sources and the cloud with an evenly distribution to improve file availability.

Our policy used the algorithm of Kernighan and Lin to partition the fog devices into two regions and the betweenness centrality index, based on the shortest path between the data sources, was used to detect the most central nodes of both regions to place two replicas of the files. A final third replica was placed in the node with the highest centrality when also the cloud provider was considered to calculate the betweenness centrality.

We simulated 22 different experiments that were varied in size from a reference scenario. The placement of our algorithm was compared with our previous proposal for single replica placement (a file placement policy without replication of the files)   and with FogStore (a previous replica-based policy proposed by Mayer et al.~\cite{DBLP:conf/fwc/MayerGSR17} and Gupta and Ramachandran~\cite{Gupta:2018:FGK:3210284.3210297}).

The results shows that our policy improved the network latency with regard to the single-file and the FogStore policies, for both reading (transmissions between the data sources and the replicas) and writing operations (transmissions between replicas and the cloud provider). On the contrary, the network usage, measured in terms of the number of transmitted packets, is higher for the replica-based policies (our proposal and the FogStore) in the case of the writing operations. This is the expected result since all the file replicas need to be updated in the writing operations and, consequently, a higher number of packets are transmitted through the network. Although that, our policy increases the network usage less than the FogStore policy. For the reading operations, our policy clearly generates a lower use of the network than the other two policies. Finally, both replica-based policies show a similar behaviour in terms of file availability, but with important benefits with regard to the single-file policy.

The first future work that emerges from this work is to optimize latency, network usage and file availability by also varying independently the number of replicas of each file. Additionally, the definition of an optimization policy for the case of splitting the files in chunks is also a second interesting future work. Finally, the use of more complex file chain flows could be modeled and studied with our proposed algorithm.

\section*{Acknowledgments}

This work was supported by the Spanish Government (Agencia Estatal de Investigaci\'on) and the European Commission (Fondo Europeo de Desarrollo Regional) through grant number TIN2017-88547-P (MINECO/AEI/FEDER, UE).

\bibliography{references}

\end{document}